           \def\OldLaTeX{0}  

\ifnum\OldLaTeX=1 
    \documentclass[12pt,letterpaper]{article}
\else
    \documentstyle[12pt]{article}
\fi

\newcommand{\r}{\vec{r}}

\newcommand{\runo}{\vec{r}_1}
\newcommand{\rk}{\vec{r}_k}

\newcommand{\rdue}{\vec{r}_2}
\newcommand{\rduep}{\vec{r}_2^{~\prime}}
\newcommand{\rnpuno}{\vec{r}_{n+1}}
\newcommand{\runop}{\vec{r}_1^{~\prime}}
\newcommand{\rnpunop}{\vec{r}_{n+1}^{~\prime}}
\newcommand{\ro}{\vec{r}_0}

\newcommand{\druno}{\dot{\vec{r}}_1}

\newcommand{\drnpuno}{\dot{\vec{r}}_{n+1}}

\newcommand{\ri}{\vec{r}_i}

\newcommand{\ra}{\vec{r}_A}
\newcommand{\rb}{\vec{r}_B}

\newcommand{\xo}{x_0}
\renewcommand{\xi}{x_i}

\newcommand{\xa}{x_A}
\newcommand{\xb}{x_B}

\newcommand{\yo}{y_0}
\newcommand{\yi}{y_i}

\newcommand{\ya}{y_A}
\newcommand{\yb}{y_B}

\newcommand{\vuno}{{\vec{v}_1}}
\newcommand{\vunop}{{\vec{v}_1^{~\prime}}}
\newcommand{\vnpuno}{{\vec{v}_{n+1}}}
\newcommand{\vnpunop}{{\vec{v}_{n+1}^{~\prime}}}
\newcommand{\vdue}{{\vec{v}_2}}

\newcommand{\uunop}{{\vec{u}_1^{~\prime}}}
\newcommand{\unpuno}{{\vec{u}_{n+1}}}
\newcommand{\unpunop}{{\vec{u}_{n+1}^{~\prime}}}

\newcommand{\buno}{{\vec{b}_1}}
\newcommand{\bunop}{{\vec{b}_1^{~\prime}}}
\newcommand{\bnpuno}{{\vec{b}_{n+1}}}
\newcommand{\bnpunop}{{\vec{b}_{n+1}^{~\prime}}}

\newcommand{\vi}{{\vec{v}_i}}

\renewcommand{\t}{\tau}
\renewcommand{\to}{t_0}
\newcommand{\tb}{\bar{t}}
\newcommand{\tbn}{\bar{t}_n}

\newcommand{\ti}{t_i}
\newcommand{\tbuno}{\bar{t}_1}
\newcommand{\tbdue}{\bar{t}_2}
\newcommand{\tbkpuno}{\bar{t}_{k+1}}
\newcommand{\tbk}{\bar{t}_k}

\newcommand{\toi}{(t_0-t_i)}
\newcommand{\tbunoto}{(\bar {t}_1+\tau -t_0)}

\newcommand{\tobn}{(t_0-\bar {t}_n)}

\newcommand{\roi}{(\vec{r}_0-\vec{r}_i)}
\newcommand{\rob}{(\vec{r}_0-\vec{r}_B)}

\renewcommand{\v}{\vert}
\renewcommand{\d}{\delta}

\newcommand{\E}{{\cal E}}

\newcommand{\g}{\gamma}

\newcommand{\p}{\prime}
\newcommand{\non}{\nonumber}
\newcommand{\rff}[1]{(\ref{#1})}

\newcommand{\pa}{\partial}
\newcommand{\lb}{\label}
\newcommand{\be}{\begin{equation}}
\newcommand{\ee}{\end{equation}}
\newcommand{\bea}{\begin{eqnarray}}
\newcommand{\eea}{\end{eqnarray}}

\begin{document}

\thispagestyle{empty}

\addtocounter{footnote}{1}
\renewcommand{\thefootnote}{\fnsymbol{footnote}}

\vspace{-20pt} 

\rightline{TIT/HEP-338/COSMO-75}\par  


\vspace{18pt} 

\begin{center}
{ {\Large {\bf
Time machines and\\ the Principle of Self-Consistency \\
as a consequence of \\ the Principle of Stationary Action (II): \\
the Cauchy problem for \\
\vskip 0.3cm
a self-interacting relativistic particle}}}

\vspace{18pt}

A. Carlini\\ 
{\small\em {Tokyo Institute of Technology, Oh-Okayama, Meguro-ku,
Tokyo 152, Japan}}\\[1.2em] 

I. D. Novikov\\ 
{\small\em {The Copenhagen University Observatory, 
Juliane Maries Vej 30, DK-2100 Copenhagen \O , Denmark}}\\ 
{\small\em{NORDITA, Blegdamsvej 17, DK-2100 Copenhagen \O ,
Denmark}} \\
{\small\em {TAC, Juliane Maries Vej 30, DK-2100 Copenhagen \O ,
Denmark}} \\
{\small\em {Astro Space Center of the P.N. Lebedev Physical
Institute, Profsoyuznaja 84/32, Moscow,}}\\
{\small\em {117810, Russia}}\\[1.2em] 

\vspace{24pt}
\end{center}

\newpage 

{\centerline{\bf Abstract}}

\bigskip
We consider the action principle to derive the classical, relativistic
motion of a self-interacting particle in a 4-D Lorentzian spacetime containing
a
wormhole and which allows the existence of closed time-like curves.
In particular, we study the case of a pointlike particle 
 subject to a `hard-sphere' self-interaction potential 
and which can traverse the wormhole an arbitrary number of times, and
show that the only possible trajectories for which the classical
action is stationary are those which are globally self-consistent.
Generically, the multiplicity of these trajectories 
(defined as the number of self-consistent solutions to the equations of
motion beginning with given Cauchy data) is finite, and it becomes infinite
if certain constraints on the same initial data are satisfied.
This confirms the previous conclusions (for a non-relativistic model) by
Echeverria, Klinkhammer and Thorne that the Cauchy initial value
problem in the presence of a wormhole `time machine' is classically 
`ill-posed' (far too many solutions).
Our results further extend the recent claim by Novikov et al.
that the `Principle of self-consistency'  is a natural
consequence of the `Principle of minimal action.'

\newpage

\addtocounter{footnote}{-\value{footnote}}
\renewcommand{\thefootnote}{\alph{footnote}}
\section{Introduction}

Since the original works by \cite{god}, there has been a long debate
over the issue whether the laws of physics might allow for the
existence of closed time-like curves (CTCs) inside our universe [2--14].
Macroscopic CTCs might be easily realized as a semiclassical consequence of
the `quantum foam' structure of spacetime at Planck scales (see, e.g.,
Refs.~\cite{haw}-\cite{whe}.).
Close to this scale, it is conjectured that spacetime might allow for
non-trivial topological fluctuations, for example in the form of
wormholes, intuitively speaking 4-d `handle-like' geometries, whose
two `mouths' join distant regions of spacetime.
The matter content which is required to produce a static wormhole
should violate the averaged weak energy condition (AWEC) [3-6],
\footnote{Recently, also dynamical wormhole
solutions which satisfy the weak and
dominant energy conditions have been found, see, e.g., \cite{dyn}.}
and other properties might be important as well (cf. Ref. \cite{gibb}).
Classically,\footnote{Just in order to avoid possible confusion, throughout 
the whole paper we will use the word `classical' as opposed to `quantum',
and `nonrelativistic' as opposed to `relativistic'.} the AWEC
 condition might be a problem, since ordinary classical energy
densities are positive.
It is not yet clear whether quantum effects
could anyway preserve the AWEC for generic cases, or
if vacuum polarization divergences in the quantum theory
can actually destabilize the wormhole
(see, e.g., Refs.~\cite{qua1,qua2}).
However, if the laws of physics actually permit the existence of traversable
wormholes, then generic relative motions of the two
wormhole's mouths, or equivalently generic gravitational redshifts at the 
mouths due to external gravitational fields, could in principle produce
CTCs looping through it [4, 7-8]: 
when traversed from mouth to mouth, the wormhole acts as
a `time machine' allowing one to travel into the past or
into the future.
For spacetimes with CTCs, past and future are no longer `globally'
distinct [9, 10-11].
In particular, as originally pointed out in Ref.~\cite{nov1},
events on CTCs should causally influence each other along the
`loops in time' in a self-adjusted, consistent way.
This requirement has been explicitly formulated
as the `Principle of self-consistency,' according to which
{\it the only solutions to the laws of physics that can occur
locally in the real universe are those which are globally
self-consistent} [7, 9, 12--14].
For further discussion and references about these and other wormhole
related topics we refer the interested reader to Ref. \cite{visserbook}.

In a recent paper \cite{us} (from now on referred to as TM-I)
we showed that the `Principle of self-consistency'
actually needs not to be imposed as an independent assumption 
which is necessary in order
to make sense of spacetimes with CTCs, but instead can be seen
as a direct consequence of the more fundamental
`Principle of minimal action.'
In particular, in TM-I we considered the simple model of a nonrelativistic
particle which is constrained
to have fixed initial and final positions, to loop through the wormhole
once and to interact with itself by means of a `hard-sphere', elastic
potential.
We used the action principle to derive the classical trajectories, and
found that the only possible solutions which minimize the action
are those which are globally self-consistent.
In the case of coplanar motion with respect to the wormhole's mouths,
the possible, globally self-consistent trajectories in which the
particle's copies collide are of three types, depending on the 
possible ways in which momentum is exchanged at the collision event.
These results led us to formulate the conjecture that
the `Principle of self-consistency' should be a consequence of the `Principle
of minimal action' for all physical phenomena.
The extension of this model to the case of 3-d motion has been also considered
in Ref. \cite{tmII} (from now on referred to as TM-II).

In a previous series of papers \cite{ekt,nov5},
a similar model for the
motion of a nonrelativistic `billiard ball'-like particle
in the spacetime containing a wormhole `time machine' was
considered in the context of a Cauchy initial value problem, where the
initial path and speed of the particle are assumed to be fixed (for
a general discussion also including the case of a scalar field, see
Ref.~\cite{con}).
In the case of elastic self-interaction 
of the particle, it was shown \cite{ekt} that generic classes
of Cauchy data with initial velocity $\vert \vi\vert >D/\tau$
(where $D$ and $\tau$ respectively denote the wormhole mouths' separation
and the time displacement into the past after the wormhole traversal) 
have multiple, and even infinite numbers of globally
self-consistent solutions to the equations of motion
(trajectories where the particle is initially at rest far from the
wormhole have, in fact, multiplicity one), with no evidence for
non self-consistent trajectories.\footnote{
The extension to the case of an inelastic self-collision of
`billiard ball'-like particles was made in Ref.~\cite{nov5}.}

The present paper is to be intended as a generalization of the 
analysis made in the previous papers TM-(I,II) 
to the case of the Cauchy initial
value problem for the motion of a `billiard ball', pointlike,
relativistic particle multiply looping 
through a wormhole `time machine' and (elastically)
self-interacting via a `hard-sphere' potential.
In particular, we address again the issues of the existence of globally
self-consistent solutions to the equations of motion in the presence of
the wormhole `time machine', of their derivation from a `Principle
of stationary action', and therefore argue against the actual need
of imposing the `Principle of self-consistency' as an extra independent
assumption in the laws of physics.
We then make further contact with the non-relativistic analysis of
Ref. \cite{ekt}, checking for the multiplicity of the globally
self-consistent solutions, and for the classical well-posedness of the
Cauchy initial value problem.

The outlay of the paper is the following.
We start in Section \ref{Sect:Model} 
by introducing the main formulas for the kinematics,
the dynamical equations and conservation laws for the Cauchy problem
in the model of a short range,
`hard-sphere' potential (effectively treating the
particle as a pointlike `billiard ball').
In Section 3 we state the main lines for the analysis of the stationary 
points of the action describing the classical motion of the particle
in three spatial dimensions, 
separately analyzing the cases without and with collisions.
In particular, we show the local solutions to the 
conservation laws of relativistic energy and momentum for the case of
coplanar motion of the particle with respect to the wormhole mouths.
Then, in Section 4, we analyze the globally self-consistent solutions
to the Cauchy initial value problem (distinguishing among the
various possibilities of the initial Cauchy data 
for the modulus and direction of
the velocity of the particle), study their multiplicity
and show (explicit and detailed formulas for the case of a coplanar
motion of the particle with respect to the wormhole's mouths
are presented in the Appendix A.2)
that the action is stationary along all these trajectories, 
therefore concluding that the `Principle of self-consistency'
is a direct consequence of the `Principle of stationary
action'.
We conclude in Section \ref{Sect:Discussion} with some discussion
and comparison with previous results \cite{con}-\cite{ekt}
presented in the context of
a similar Cauchy initial value 
problem for the motion of a non-relativistic particle
in the presence of a wormhole `time machine'.

\section{The model}
\label{Sect:Model}

We consider the relativistic motion of a self-interacting particle of mass $m$
in the background with a wormhole `time machine' and
 with a fixed set of initial
Cauchy data.
Similarly as done in Refs. TM-(I,II), the mouths of the wormhole are here 
treated as pointlike and to be infinitely heavy, so that we
can neglect the recoil effect on the geometry when the particle
traverses the wormhole.
In particular, we suppose that the mouths are at rest in some reference
frame, and consider the problem using this frame.
Spacetime outside the `time machine' is approximated to
be Minkowskian.
Our discussion is essentially independent of other features
defining the internal structure of the wormhole (although it is
consistent, e.g., with the choice of `traversal rules' suggested
in Ref.~\cite{ekt}).
Finally, we depart from Refs. TM-(I,II) by allowing
for motions in which the particle can
traverse the wormhole an arbitrary number ($n$) of times.

The motion can be schematically described in the following way.
The particle is assumed to start at time $t_i$ in the position $\ri$,
with velocity $\vi$, and to enter the first mouth (B) of the wormhole at 
time $\tbuno+\t$ (position $\rb$).
Then it exits from the other mouth (A) at the earlier time $\tbuno$
(position $\ra$) and moves towards mouth B, where it enters at time
$\tbdue +\tau$, again exits from mouth A at time $\tbdue$ and so on,
 making $n$ of such wormhole traversals (i.e., exiting
mouth A at time $\tbk$ and reentering mouth B at time $\tbkpuno +\t$, for
$k=1, ... n-1$).
The particle definitively leaves the wormhole from mouth A at time $\tbn$.
For the particle itself (in its proper time), each motion through the
wormhole happens almost istantaneously, as the path length of the
wormhole handle is assumed to be infinitely short.
According to an external observer, instead, after each of the traversals
of the `time machine' the particle travels back in time by the amount 
$\Delta t=-\t$, where by definition $\t>0$.\footnote{These boundary conditions
will be `relaxed' only in Section 4.1.c and Appendix A.2.c, where we will 
consider the motion of a particle initially heading towards mouth A
(with Cauchy data $\ri, \vi$ at $\ti$), entering it at time $\tbn$ and,
after making $n$ wormhole traversals (exiting mouth B at time $\tbk +\t$
and reentering mouth A at time $\bar t_{k-1}$, for $k=2, ... n$), finally
exiting from mouth B at time $\tbuno +\t$.
In this case, after each wormhole traversal, the particle travels
forward in time by the amount $\Delta t= +\t$.}
 
We first analyze the generic motion of the particle
in three spatial dimensions, and
then discuss the detailed trajectories for the case of
coplanar motion with respect to the wormhole's mouths.

The conditions for the wormhole to act as a `time machine' during each 
of its traversal and also after a multiple number $n$ of traversals
are, respectively,
\begin{equation}
\left. \begin{array}{l}
\t > \v \rb-\ra\v/c, \\
\tbuno +\tau >\tbn,
       \end{array}
\right.
\label{duea}
\end{equation}
where $c$ is the speed of light. 

In the case of a general potential $V$ for the self-interaction of 
the particle, and depending on the relations between the times
$\tbk, \tbkpuno +\tau$ ($k=1, ... n-1$) and the times $\ti$, $\tbuno+\t$ 
and $\tbn$,
at each time $t$, in principle, one might have from a minimum of one to 
a maximum of $n+1$  copies of the relativistic particle existing at the 
same time, copies which are treated as independent objects having 
multiple, self-interactions between themselves and with the wormhole mouths.
If the particle enters mouth A (at time $\tbn$) and finally appears from mouth
B in the future (at time $\tbuno +\tau$), at any time $t$ there is at
most one copy of the particle in the space external to the wormhole.
In particular, we can describe  the trajectories of the copies of the particle
in the following way: $\runo$ for the copy moving from the initial 
position $\ri$ to the first wormhole mouth entrance (position $\rb$),
$\rk$ ($k=2, ... n$) for the copies performing the multiple wormhole passages
(between wormhole mouth's positions $\ra$ and $\rb$), and finally 
$\rnpuno$ for the last copy definitely leaving the second mouth (position
$\ra$).

We then assume the Cauchy initial data for the relativistic particle, 
\begin{equation}
\left. \begin{array}{l}
\runo(t_i)\equiv \ri, \\ 
\druno(t_i)\equiv \vi
       \end{array}
\right.
\label{unoa}
\end{equation}
(initial position and velocity), to be fixed in a region devoid
of CTCs (cf., e.g., Ref. \cite{ekt}, where the initial path and velocity
are fixed instead). 

The time displacement $\t$ in the wormhole, as well as the positions
of the wormhole's mouths
are also assumed to be known.
The wormhole entrance and exit conditions on the position of the 
particle are
formally summarized as the constraints\footnote{In Section 4.1.c and Appendix
A.2.c, for the case of collisionless motion of a particle traversing
the wormhole from mouth A to mouth B and travelling forward in time,
we will also consider the modified boundary conditions $\vec{r}_{n-k+1}(\tbk)
=\ra$ and $\vec{r}_{n-k+2}(\tbk +\t)=\rb$, for $k=1, ... n$.}
\be
\left. \begin{array}{l}
\vec{r}_{k+1}(\bar t_{k})\equiv \ra\;\; ; \;\; k=1, ... n,
\\[0.5em]
\vec{r}_k(\bar t_k+\t)\equiv \rb\;\; ; \;\; k=1, ... n.
      \end{array}
\right.
\lb{unob}
\ee

For simplicity, we also choose to work in the ansatz
\footnote{Actually, since the
relation between the times $t_i$ and $\tbn$ turns out not to be relevant
for the kinematics of the particle, the choice \rff{ansatz} is made
mainly for illustrative purposes.} 
\be
\tbn >t_i. 
\label{ansatz}
\ee

\subsection{`Billiard ball' pointlike particle}
\label{Sect:HardSphere}

In principle it would be possible to write down the classical action 
describing a general interaction among the particle copies and the wormhole 
mouths.\footnote{For instance, one can assume that both the particle and the
wormhole are `dressed' with electromagnetic charge.
In this case one should take into account
the interactions between the particle copies and the wormhole mouths -
when a charged particle enters or exits the wormhole the charge of
the wormhole mouths also changes \cite{fro}. 
The charged particles will then have, in general, non zero acceleration 
and emit radiation.
Radiation is also expected to propagate through the wormhole throat.
Finally, due to relativistic effects, all interactions should be described in
terms of `retarded' times.}
However, as in general the problem of solving the equations of motion by
minimizing such an action turns out not to be straightforward (see also 
Ref. TM-I), we will not address it here.

Instead, the nature of the trajectories can be greatly simplified
assuming that the particle and wormhole mouths are
 not charged and by working with the model of a short 
range, `hard-sphere' self-interaction potential for the particle


\begin{equation} 
V(r)=\hat V\theta (r_s-r);\ \ 
\hat V\rightarrow\infty,\ \
r_s\rightarrow 0,
\label{2}
\end{equation} 
where we have introduced the variable
\begin{equation}
\r(t)\equiv \rnpuno(t)-\runo(t)\;\; ; \;\; r\equiv |\vec{r}|.
\label{10}
\end{equation}
In this model, the particle behaves
like a (small)\ `billiard ball' and 
we essentially neglect the interaction of the
copies of the particle along most of 
their motion in Region~II, except at
the point of the (eventual) elastic collision.
The effect of the potential is limited to an infinitesimally small period
of time around $\to$, the time of the (eventual) collision.
We want to make clear that, since treating the particle like a 
`billiard ball' instantaneously interacting with itself under the action of
a `hard sphere' potential would implicitly assume that the velocity
of sound is infinite, the choice of the ansatz \rff{2} can be made
consistent with a relativistic description of the motion only
if the particle is treated as {\it pointlike}.
In this ansatz one can then also exclude (neglecting relativistic `retarded' 
effects in the potential) eventual interactions among copies $k=2, ... n$
of the particle and among these and the other two copies $k=1, n+1$,
as this only requires certain restrictions
on the time displacement produced by the wormhole (see Eq. \rff{nc} in 
Section 3.1) and
on the geometry of the trajectories in Region II (see Section 3.3.1).
As a consequence of this choice, the actual ordering
between the times $\tbk, \tbkpuno +\t$ ($k=1, ... n-1$) and $\ti, \tbn,
\tbuno+\t$ becomes
essentially irrelevant for the purpose of our analysis. 
We exclude the possibilities of
more than one collision between the two copies of the particle in Region II,
and also 
Cauchy data such that the particle is initially moving along the direction
of the line connecting the two wormhole mouths. 

In fact, proceeding along lines similar to Refs. TM-(I,II), it seems convenient
to describe the motion of the relativistic particle in the
presence of the `time machine' using 
the following four main separate regions:\par 
I) $t_i<t<\tbn$~: only the first copy of the particle 
with position $\runo(t)$
is present;\par
II) $\tbn<t<\tbuno+\t$~: two copies of the particle with positions 
$\runo(t)$ and 
$\rnpuno(t)$, interacting via the potential $V$, are present;\par 
III) $t>\tbuno+\t$~: only the $(n+1)$-th copy of the particle with
position
$\rnpuno(t)$ is present;\par
IV) $\tbk<t<\tbkpuno +\t$~, $k=1, ... n-1$: 
from one to $n-1$, non self-interacting copies with position $\rk(t)$ 
($k=1, ... n-1$), moving between mouth A and 
mouth B are present (regardless of the eventual
presence of one or both of the $k=1, n+1$ copies of the particle in Regions
I-III).

The total action describing such a motion
is the sum of the actions of the single paths in each separate
region (subject to obvious continuity conditions for the position
of the copies of the particle at times $\tbk$ and $\tbk+\t$,
$k = 1, ... n$), i.e., 
\begin{eqnarray}
S&=&-mc^2\left \{\int^{\tbn}_{t_i}d\tilde t~[\gamma_1(\tilde t)]^{-1}+
\int^{t}_{\tbuno+\t}d\tilde t~[\gamma_{n+1}(\tilde t)]^{-1}
+\sum_{k=1}^{n-1}\int^{\bar t_{k+1}+\t}_{\bar t_k}d\tilde t~
[\gamma_{k+1}(\tilde t)]^{-1}\right \}
\non \\
& &{} -\int^{\tbuno+\t}_{\tbn}d\tilde t\biggl\{mc^2\left 
[[\gamma_1(\tilde t)]^{-1}+
[\gamma_{n+1}(\tilde t)]^{-1}\right ]+ V(r(\tilde t))\biggr\}
\non \\
&\equiv& S_1(t_i, \tbn)+S_{n+1}(\tbuno+\t, t)+S_{\Sigma}+
S_{1, n+1}(\tbn, \tbuno+\t).
\label{6}
\end{eqnarray}
The general procedure consists in imposing the principle of
stationarity of the action, considering continuous paths for which
the initial and final coordinates of the particle (respectively, at times $\ti$
and $t$) are (provisorily) held fixed. 
One then obtains the classical equations of motion in each
of the four Regions~I, II, III and IV, which can be solved separately
subject to the Cauchy initial conditions \rff{unoa}.\footnote{Note that the 
Euler-Lagrange equations, although formally derived from Eq. \rff{6} by
a variational principle which assumes fixed initial and final particle 
positions, are second order differential equations which can be solved
by giving the initial coordinate and velocity of the particle.
It is in this sense that one can consider the Cauchy initial value problem
for the classical trajectories by starting from the action principle,
see Ref. \cite{landau}.}

\paragraph{Regions I and III.}
By the variation of the action $S_1$ in Eq.~\rff{6} with respect to $\runo$ in
the first region, and of $S_{n+1}$ with respect to $\rnpuno$ in the third 
region,
we find
\begin{equation}
\left. \begin{array}{lcl}
{\d S_1\over\d\runo}=0 \;\; & \Rightarrow &\;\;  (\gamma_1\druno)^{\cdot}=0, 
\\[0.5em] 
{\d S_{n+1}\over\d\rnpuno}=0 \;\; & \Rightarrow & \;\;  
(\g_{n+1}\drnpuno)^{\cdot}=0. 
       \end{array}
\right.
\label{7}
\end{equation}
Equations \rff{7} clearly represent linear motion.

\paragraph{Region II.}
By varying the action $S_{1, n+1}$ in Eq. \rff{6}
with respect to $\runo$ and $\rnpuno$,
we have the following equations of motion
\begin{equation}
\left. \begin{array}{lcl}
{\d S_{1, n+1}\over\d\runo}=0\;\; & \Rightarrow & \;\; 
mc^2(\gamma_1\druno)^{\cdot}=V^{\p}(r){\r\over r}, 
\\[0.5em] 
{\d S_{1, n+1}\over\d\rnpuno}=0 \;\; & \Rightarrow& \;\; 
mc^2(\g_{n+1}\drnpuno)^{\cdot}=-V^{\p}(r){\r\over r}. 
       \end{array}
\right.
\label{9}
\end{equation}
If we further introduce the relativistic momenta
\be
{\vec{p}}_k\equiv m \g_k\dot{\vec{r}}_k\;\; ;\;\; k=1, n+1,
\lb{11}
\ee
then, summing Eqs.~\rff{9} we obtain
\begin{equation}
[\dot{\vec{p}}_1+\dot{\vec{p}}_{n+1}](t)=0. 
\label{12}
\end{equation}

Moreover, noting that the action \rff{6} is invariant with respect to time
translations, we can make use of the variational principle to show
that (see, e.g., Ref. \cite{landau})  
\be
\sum_{i=1, n+1}\left [\dot{\vec{r}_i}
{\d S_{1, n+1}\over \d \dot{\vec{r}_i}}-{\cal L}_{1, n+1}\right ]=\E
\lb{13}
\ee
(where $\E = {\mbox{\em const}}$),
from which we obtain the first integral for the relativistic motion
\begin{equation}
mc^2\g_1(t)+mc^2\g_{n+1}(t)+ V[r(t)]=\E.
\label{13a}
\end{equation}
Equations~\rff{12} and \rff{13a} are nothing but the relativistic laws of 
momentum and energy conservation for the motion of the two copies of the
particle in Region~II.\footnote{Taking the potential \rff{2}, it is clear 
that Eq.~\rff{13a} is not well defined at the point $r=r_s$.
However, it is possible to show 
(similarly as done in Ref. TM-I, see Appendix~\ref{App:Hard}) 
that the total relativistic  energy $m(\gamma_1 +\gamma_{n+1})c^2$ is 
conserved before and after the collision.}

Moreover, for $r>r_s\sim 0$ we have  $V(r)=0$, and
Eqs.~\rff{12} and \rff{13a} 
state that, everywhere in Region~II except at the point
of eventual collision, the motion of the two copies of the particle
is also linear.

\paragraph{Region IV.}
By the variation of the action $S_{\Sigma}$ in Eq. \rff{6}
with respect to $\rk$ we
get the following equations of motion

\be
{\d S_{\Sigma}\over\d\vec{r}_k}=0\;\; \Rightarrow \;\; 
(\gamma_k\dot{\vec{r}_k})^{\cdot}=0\;\; ; \;\; k=2, ... n. 
\lb{14}
\ee
Eqs. \rff{14} clearly imply that also the motion in Region IV is linear for
each of the copies $k=2, ... n$ of the relativistic particle.
In details, in terms of the relativistic $\gamma$-factors we have that
\be
\g_k(\bar t_{k-1})=\g_k(\bar t_k+\t)=const \;\; ; \;\; k=2, ... n. 
\lb{18}
\ee

Finally, taking the variation of the action \rff{6} with respect to
$\tbk$, and excluding the possibility of collisions on the verge of the
wormhole's mouths (in other words, assuming that
$V[r(\tbk)]=V[r(\tbk +\t)]=0$, $k=1, ... n$), 
we get the set of conditions\footnote{Anticipating the discussion and 
notation of Section 3.3, in Eqs. \rff{20}-\rff{21}
we introduce a prime to distinguish  the velocity of the first
copy of the particle before its first entrance into the wormhole mouth
B and after the eventual collision with its $(n+1)$-th copy (finally
exiting from mouth A), from its - unprimed - velocity before the same eventual
collision.}
\be
{\pa S\over \pa \bar t_k}=0\;\; \Rightarrow \;\; \g_k(\bar t_k+\t)=\g_{k+1}
(\bar t_k)=\g_1^{~\prime}(\tbuno+\t)=\g_2(\tbuno)\;\; ; \;\; k=2, ... n,
\lb{20}
\ee 
which, on use of Eqs. \rff{18}, finally gives
\begin{equation} 
\v\vec{v}_k\v=\v\vec{v}_1^{~\prime}\v=
\v\vec{v}_{n+1}\v\;\; ; \;\; k=2, ... n,
\lb{21}
\end{equation}
stating that the energy of the particle at each entrance and exit from the
wormhole mouths must be conserved.


In conclusion, for the motion of a relativistic
particle constrained to traverse the wormhole an arbitrary number ($n$) of
times and to have a given set of fixed initial Cauchy data,
we have to distinguish between the two cases:

\paragraph{i) trajectories without self-collision.}
In this case, the first copy of the particle moves linearly from the
initial position $\ri$ (and velocity $\vi$)
at time $t_i$ until it enters the wormhole
mouth B at time $\tbuno+\t$.
Similarly, after the $n$ wormhole traversals, the $(n+1)$-th 
copy of the particle moves linearly
from  mouth A at time $\tbn$ along a certain direction which
is specified by the internal wormhole geometry.\footnote{As already remarked
at the beginning of Section 2, in Section 4.1.c and Appendix A.2.c we will
also consider the case of a particle linearly moving from $\ri$ 
(velocity $\vi$) at time $\ti$, to mouth A at time $\tbn$, and then finally
exiting (after $n$ wormhole traversals) and linearly moving from mouth B
at time $\tbuno +\t$.}

\paragraph{ii) trajectories with self-collision.}
In this case, instead, the motion for the first ($(n+1)$-th) copy 
of the particle
is linear from the initial position $\ri$ at time $t_i$
(from the wormhole mouth A at time $\tbn$) up to the collision event, with
coordinates

\begin{equation} 
\runo(\to)=\rnpuno(\to)\equiv\ro. 
\lb{311b}
\end{equation} 
After the collision, the motion of the first copy of the particle
is again linear up to the wormhole mouth B at time $\tbuno +\t$,
and also the $(n+1)$-th copy moves linearly.
Of course, the directions of the trajectories for the first ($(n+1)$-th) copy
of the particle will be, in general, different before and after the
collision (see Sections~\ref{Sect:Sub:WSC} and 3.3.1).

In the case of a short-range potential, therefore, the stationarity
problem is simplified as
the trajectories will depend only on the parameters
\begin{equation} 
\tbuno,\ \ \tbn,\ \ \to,\ \ \ro
\label{30}
\end{equation} 
and, of course, on the initial Cauchy data $\ri, \ \ \vi, \ \ \ti$.

The problem is now to look for the stationary points (if any)\ 
of the action
\rff{6}, evaluated along the classical 
trajectories \rff{7}, \rff{9} 
and \rff{14}, with respect to the 
parameters \rff{30}.

\section{The trajectories}
\subsection{Motion in Region IV} 
On account of our assumptions about at most a single self-collision
of the particle in Region II, the analysis of the motion in Region IV
is necessary mainly in order to determine the relationship between
the times of first entrance and last exit of the particle itself from 
the wormhole. 

Solving explicitly for the trajectories in Region IV, i.e. 
Eqs. \rff{14}-\rff{18}, subject to the conditions \rff{21} and to the 
boundary conditions \rff{unob}, we get
\be
\vec{r}_k(t)=\v\vec{v}_1^{~\prime}\v{(\rb-\ra)\over \v \rb
-\ra\v}(t-\bar t_{k-1})+\ra\;\; ; \;\; k=2, ... n.
\lb{24}
\ee
In particular, the times of entrance and exit, respectively,
from the wormhole mouths
B and A, are related by the obvious recurrence relations
\be
\bar t_k+\t=\bar t_{k-1}+{\v \rb-\ra\v\over \v \vec{v}_1^{~\prime}\v}
\;\; ; \;\; k=2, ... n.
\lb{26}
\ee
Iterating Eq. \rff{26} for $k=2, ... n$, we can finally get the relation
between the time of first entrance of the particle into the wormhole
mouth B ($\tbuno +\t$) and the time of last exit from
mouth A ($\tbn$) as 
\be
\bar t_n=\bar t_1+(n-1)\left [{\v \rb-\ra\v\over \v \vec{v}_1^{~\prime}\v}-\t
\right ].
\lb{27}
\ee
From inspection of Eq. \rff{27}, it is straightforward to check that the
condition of the wormhole acting as a `time machine' (i.e., that $\tbuno 
+\t> \tbn$, see the last of Eqs. \rff{duea}) even {\it after} its multiple 
traversals ($n>1$) by the 
particle turns out as\footnote{For the case of a particle first entering 
mouth A, traversing the wormhole $n$ times and finally exiting mouth B without
any self-collision (Section 4.1.c and Appendix A.2.c), Eqs. \rff{24}-\rff{27}
should be replaced by:
$\vec{r}_k(t)=\v\vi\v(\ra-\rb)(t-\bar t_{n-k+2}-\t)/\v\rb-\ra\v +\rb$,
$\bar t_k+\t=\bar t_{k-1}-\v\rb-\ra\v/\v\vi\v$ (for $k=2, ... n$) and
$\bar t_1=\bar t_n+(n-1)[\v\rb-\ra\v/\v\vi\v +\t]$, for which of course
the condition $\tbuno +\t>\tbn$ is trivially satisfied.}
\be
\t >\left ({n-1\over n}\right ){\v \rb-\ra\v\over \v \vec{v}_1^{~\prime}\v}.
\lb{28}
\ee

Finally, from Eq. \rff{26} we can also easily get the condition for no 
self-collision between the $k=2, ... n$ copies of the relativistic
particle in Region IV, i.e. 
\be
\t\not = {\v \rb-\ra\v\over \v \vunop\v}.
\lb{nc}
\ee
  
\subsection{Motion in Region II: no collision}
\label{Sect:Sub:NSC} 

We first consider the variational problem in Region II for the
case of `no collision' (i.e., when the separation $r$ between the
two copies of the particle is always greater than $r_s$).
In this case the trajectories can be 
parametrized as 
\begin{equation}
\left. \begin{array}{ll}
\runo(t),\ \ &t_i<t<\tbuno +\t, 
\\
\rnpuno(t),\ \ & t>\tbn,  
       \end{array}
\right.
\lb{34}
\end{equation}
with the (constant) velocities subject to the conditions (see Eq. \rff{9}
for $V=0$ and Eq. \rff{21})
\begin{equation}
\left. \begin{array}{ll}
\v\vuno\v=\v\vnpuno\v =\v\vi\v ,\ \ &\vnpuno\equiv \v\vi\v \vec{k}_{n+1} 
       \end{array}
\right.
\lb{35}
\end{equation}
(with $\vec{k}_{n+1}$ a unit vector).
Then, the solutions
of the equations of motion \rff{9},
subject to
the Cauchy data \rff{unoa} and the boundary conditions \rff{unob}, are given
by the linear trajectories
\begin{equation}
\left. \begin{array}{l}
\runo(t; \tbuno)=\vi (t-\tbuno-\t)+\rb,
\\[0.5em] 
\rnpuno(t; \tbuno)=\v\vi\v\vec{k}_{n+1} (t-\tbn)+\ra,
       \end{array}
\right.
\lb{387}
\end{equation}
where $\tbn$ is a function of $\tbuno$ via Eq. \rff{27}.\footnote{For the
collisionless motion of a particle first entering mouth A and finally
exiting from mouth B, see Section 4.1.c and Appendix A.2.c.}

The variational problem is now extremely simple.
In fact, the action
\rff{6} evaluated along the classical trajectories \rff{387}
is a function only of the (fixed) Cauchy data,\footnote{And, of course, 
also of the direction $\vec{k}_{n+1}$ depending on the wormhole
internal geometry, which however we leave as arbitrary here.}
i.e. 
\bea
S_{cl, n-c}&=&-mc^2\left [[\gamma_i]^{-1}
\int_{\ti}^{\tbuno +\t}d\tilde t +[\gamma_{n+1}]^{-1}
\int_{\tbn}^{t}d\tilde t\right ]+S_{\Sigma, cl}
\non \\
&=&-mc^2[\gamma_i]^{-1}[t-\ti+n\t]
\lb{387a}
\eea
(where, as in the non-relativistic case discussed in Ref. TM-I,
 the contribution from the potential term is zero, see Appendix A.1)
and is stationary.
The only `unknown' in the problem is $\tbuno$, which can be fixed by 
further imposing the boundary condition \rff{unob} on the stationary
trajectories \rff{387} (see Appendix A.2).\footnote{In this and the following
Section, there is still an apparent dependence of the classical action on the
final time $t$, see Eqs. \rff{387a} and \rff{39}.
However, since the action evaluated along the trajectories which are solutions
of the Euler-Lagrange equations is already stationary at $t$ 
(see the discussion at pag. 8), $t$ is no longer a parameter of the theory
with respect to which the classical action is to be varied.}

\subsection{Motion in Region II: self-collision}
\label{Sect:Sub:WSC} 

In the case of self-collision under the action of the `hard-sphere'
potential \rff{2}, as we already remarked in 
Section 3,
the motion is also linear everywhere except at the event of the impact.
Besides the general conditions \rff{duea}, in order that the motion with
self-collision in Region II is self-consistent, we must impose the further,
obvious, time ordering conditions
\begin{equation}
\left. \begin{array}{l}
\to >\tbn, \\
\tbuno +\tau >\to, \\
\to >\ti.
       \end{array}
\right.
\label{dueaa}
\end{equation}

To clearly identify the trajectories before and after the collision,
it is then convenient to use the following notation (see Ref. TM-I) 
\begin{equation}
\left. \begin{array}{ll}
\runo(t),\ \ &t_i<t<\to, 
\\
\runop(t),\ \ & \to<t<\tbuno+\tau,  
       \end{array}
\right.
\lb{newnot1}
\end{equation}
for the first copy of the particle, and
\begin{equation}
\left. \begin{array}{ll}
\rnpuno(t),\ \ &\tbn<t<\to, 
\\
\rnpunop(t), &t>\to, 
       \end{array}
\right.
\lb{newnot2}
\end{equation}
for the $(n+1)$-th copy of the particle, with $\tbn$ to be seen as a function
of $\tbuno$ through Eq. \rff{27}, as before.

It is easy to show that the solutions of the classical equations of motion
\rff{9}, subject to the Cauchy initial data 
\rff{unoa} and to the boundary conditions \rff{unob}, are given by
\begin{equation}
\left. \begin{array}{l}
\runo(t; \tbuno, \to, \ro)=\ro +\vi(t-\to),
\\[0.5em] 
\rnpuno(t; \tbuno, \to, \ro)=\ro +(\ro-\ra)(t-\to)/\tobn, 
\\[0.5em] 
\runop(t; \tbuno, \to, \ro)=\ro +(\rb-\ro)(t-\to)/\tbunoto, 
\\[0.5em] 
\rnpunop(t; \tbuno, \to, \ro)=\ro +\vnpunop(t-\to).
       \end{array}
\right. 
\label{314}
\end{equation}
Then, with the new notation \rff{newnot1}--\rff{newnot2}, the total action 
$S_{cl, c}$ for the collision case, evaluated
along the classical trajectories \rff{314},
(again the contribution of the potential term is zero)
more simply reads
\bea
S_{cl, c}(\to, \tbuno, \ro)&=&-mc^2\biggl[[\g_i]^{-1}\int^{\to}_{t_i}
d\tilde t
+[\g_1^{~\prime}]^{-1}\int^{\tbuno+\t}_{\to}
d\tilde t
\non \\
&&+[\g_{n+1}]^{-1}\int^{\to}_{\tbn}
d\tilde t+[\g_{n+1}^{~\prime}]^{-1}\int^{t}_{\to}d\tilde t \biggr ]
+S_{\Sigma, cl}
\non \\
&= & -mc\biggl [\sqrt{\toi^2c^2-\roi^2}
\non \\
&& {} +\sqrt{\tbunoto^2c^2-\rob^2}[n\t/\tbunoto]
\non \\
& &{}+\sqrt{(t-\to)^2c^2-(\r(t)-\ro)^2}\biggr ].
\label{39}
\end{eqnarray}
The variational problem for the collision case consists in looking for
the stationary points of the action \rff{39} with respect to the
parameters $\ro$, $\to$, and $\tbuno$, i.e. 
\be
{\pa S_{cl, c}\over \pa \ro}=0\;\; ; \;\; {\pa S_{cl, c}\over \pa \to}=0\;\;
; \;\; {\pa S_{cl, c}\over \pa \tbuno}=0.
\lb{40}
\ee
From Eqs. \rff{21} and \rff{314}-\rff{40} we find the following conditions
\begin{equation}
\left. \begin{array}{l}
\vec{u}_i+\vec{u}_{n+1} =\vec{u}_1^{~\prime}+\vec{u}_{n+1}^{~\prime}, 
\\[0.5em] 
\g_i+\g_{n+1}=\g_1^{~\prime}+\g_{n+1}^{~\prime},
\\[0.5em] 
\g_1^{~\prime}=\g_{n+1},
       \end{array}
\right.
\lb{42}
\end{equation}
where we have introduced the relativistic velocities
\be
{\vec{u}}_k^{(\prime)}\equiv \g_k{\vec{v}}_k^{(\prime)}\;\; ;\;\; k=1, n+1.
\lb{41}
\ee

Eqs. \rff{42} respectively represent the conservation laws for
relativistic momentum and  energy during the collision, and the conservation of
relativistic energy at the first entrance and last exit of the particle from 
the two wormhole's
mouths.

Equations \rff{42} can in principle 
be solved either directly in the $\ro$, $\to$, and $\tbuno$
variables, or in terms of the velocity variables (for instance, in 2-d,
considering $\vunop$ and $\vnpunop$ as unknowns and $\vi$ and $\vnpuno$ as
parameters).

Using velocities as our unknowns and introducing the quantities
(see Refs. TM-(I,II))
\begin{equation}
\left. \begin{array}{l} 
\vec{a}\equiv\vec{u}_i-\unpunop =\uunop-\unpuno, 
\\[0.5em] 
\vec{b}\equiv\uunop+\unpuno, 
\\[0.5em] 
\vec{c}\equiv\vec{u}_i+\unpunop, 
       \end{array}
\right.
\lb{43}
\end{equation}
Eqs.~\rff{42} can be easily transformed into the equivalent system
of conditions
\begin{equation}
\left. \begin{array}{l}
\vec{a}\cdot\vec{b}=0,
\\[0.5em]
\vec{a}\cdot\vec{c}=0.
       \end{array} 
\right.
\lb{44}
\end{equation}
For the motion in three spatial dimensions, the most general solution
of the conservation laws \rff{42} is thus given by
\begin{equation}
\left. \begin{array}{l}
\vec{u}_i={1\over 2}(\vec{c}+\vec{a}), 
\\[0.5em] 
\unpuno={1\over 2}(\vec{b}-\vec{a}), 
\\[0.5em] 
\uunop={1\over 2}(\vec{b}+\vec{a}), 
\\[0.5em] 
\unpunop={1\over 2}(\vec{c}-\vec{a}), 
      \end{array}
\right. 
\lb{319c}
\end{equation}
for any {\em arbitrary\/} $\vec{a}$ which is {\em orthogonal\/} to 
{\em arbitrary\/} $\vec{b}$ and $\vec{c}$.
Then, in principle, in the case of a generic three dimensional 
motion, Eqs.~\rff{44} (with $\vec{a}$ given by the first of 
Eqs.~\rff{43})\ can be solved, 
using Eqs.~\rff{314}, for $\ro$, $\to$, and $\tb$ 
and therefore for the complete trajectories (see Ref. TM-II).
We shall not do that here, but only consider the simpler case of two
dimensional spatial motion.

\subsubsection{Coplanar motion}

The solutions of the conservation Eqs.~\rff{42} for the case of
two dimensional, relativistic 
motion of the copies of the particle coplanar with
respect to the wormhole mouths can be deduced from the generic three
dimensional solutions \rff{319c} by restricting to the following
ansatz for $\vec{a}, \vec{b}$ and $\vec{c}$ (see Refs. TM-(I,II))
\begin{equation}
{\mbox{\em i)}}\ \ \vec{c}=\epsilon\vec{b};\ \ {\mbox{or}}\ \ 
{\mbox{\em ii)}}\ \ \vec{a}=\vec{0},
\lb{319d}
\end{equation}
where $\epsilon$ is an arbitrary constant.

\paragraph{a) Cauchy data with generic relativistic velocity: $\gamma_i
\sim finite$}
\hfill\break
\vskip 0.3cm
The ansatz {\em i)\/} of Eq.~\rff{319d} (for $\epsilon \neq -1$)
corresponds to the case of\footnote{For this classification of the 
classical trajectories see Refs. \cite{ekt} and TM-(I,II).}
\subparagraph{$\bullet$ `Mirror exchange of velocities' rule:}
\begin{equation}
\left. \begin{array}{l} 
(u_1^{~\prime})_x=(u_{n+1})_x,
\\[0.5em] 
(u_{n+1}^{~\prime})_x=(u_i)_x,
\\[0.5em] 
(u_1^{~\prime})_y=-(u_{n+1})_y,
\\[0.5em] 
(u_{n+1}^{~\prime})_y=-(u_i)_y,
      \end{array}
\right.
\label{49}
\end{equation}
expressed in components in the frame whose $x$-axis is
along the direction of $\vec{u}_i+\vec{u}_{n+1} =\vec{u}_1^{~\prime}+
\vec{u}_{n+1}^{~\prime}$, i.e. where
\be
(u_i)_y+(u_{n+1})_y=0.
\lb{48}
\ee
In terms of velocities, the `mirror exchange' solution can also
be written as 
\begin{equation}
\left. \begin{array}{l} 
(v_1^{~\prime})_x=(v_{n+1})_x,
\\[0.5em] 
(v_{n+1}^{~\prime})_x=(v_i)_x,
\\[0.5em] 
(v_1^{~\prime})_y=-(v_{n+1})_y,
\\[0.5em] 
(v_{n+1}^{~\prime})_y=-(v_i)_y,
\\[0.5em]
\g_i(v_i)_y+\g_{n+1}(v_{n+1})_y=0. 
      \end{array}
\right.
\label{50}
\end{equation}
It is quite easy to check that this kind of trajectories is possible only 
when the relativistic particle is initially (at time $\ti$)
pointing away from both wormhole mouths.

For the particular value $\epsilon=-1$, the ansatz 
{\em i)\/} of Eq.~\rff{319d} 
no longer corresponds to the `topology' of the `mirror exchange
solutions,' but to the case of

\paragraph{$\bullet$ `Collinear velocities' rule:}
\begin{equation}
\left. \begin{array}{l}
\unpuno=-\vec{u}_i,
\\[0.5em] 
\uunop =-\unpunop,
\\[0.5em] 
\v\unpuno\v~=~\v\uunop\v~=~\v\unpunop\v=~\v\vec{u}_i\v,
       \end{array}
\right.
\lb{46}
\end{equation}
which is also equivalent, in terms of velocities, to
\begin{equation}
\left. \begin{array}{l}
\vnpuno=-\vi,
\\[0.5em] 
\vunop =-\vnpunop,
\\[0.5em] 
\v\vnpuno\v~=~\v\vunop\v~=~\v\vnpunop\v=\v\vi\v. 
       \end{array}
\right.
\lb{47}
\end{equation}
These solutions are `degenerate' in the sense that the velocities
$\vi$ and $\vnpuno$ must be along the direction identified by $\ri$ and
$\ra$ (i.e. the velocity of the first copy of the particle must
be initially pointing towards the wormhole mouth A), and similarly
the velocities $\vunop$ and $\vnpunop$ must be along the direction
identified by $\ro$ and $\rb$ (i.e. the velocity of the second copy of
the particle after the collision must be outwards pointing from
the wormhole mouth B).\footnote{The solution $\vec{c}=-\vec{b},
\vec{a}=\vec{0}$ is a `doubly degenerate' case, as it corresponds
to one dimensional spatial motion of the copies of the particle
along the line connecting the wormhole's mouths, and is not considered
here.}

The ansatz {\em ii)} of Eq.~\rff{319d} corresponds, instead, to the case of

\paragraph{$\bullet$ `Velocity exchange' rule:}

\begin{equation}
\left. \begin{array}{l} 
\uunop=\unpuno,
\\[0.5em] 
\unpunop=\vec{u}_i,
       \end{array}
\right.  
\label{51}
\end{equation}
or, in terms of velocities,
\begin{equation}
\left. \begin{array}{l} 
\vunop=\vnpuno,
\\[0.5em] 
\vnpunop=\vi.
       \end{array}
\right.  
\label{52}
\end{equation}

However, from a simple analysis of the `topology' of the motion (for
a motion with `velocity exchange'  the
trajectories $\runop$ and $\rnpuno$ should lie along the direction
of the line connecting the two wormhole mouths, i.e. parallel to
the trajectories $\rk$, $k=2, ... n$), it is easy to show that
the condition of admitting only one self-collision of the particle in Regions
I-IV requires $n=1$, i.e. only one wormhole traversal is allowed in
this case.

Finally, there is also the `trivial' solution in which the
velocities of the copies of the particle do not change before and
after the collision.
This actually corresponds to the `no collision'
 case which we already considered
in Section \ref{Sect:Sub:NSC}. 

\paragraph{b) Cauchy data with ultrarelativistic velocity I: $\gamma_i
\rightarrow \infty ~;~~\gamma_k\sim finite ~,~~k\not =i$}
\hfill\break

In the case of initial Cauchy data with ultrarelativistic velocity
(but the velocities of all other copies of the particle only relativistic,
i.e. with $\gamma_k\sim finite$ for $k\not = i$), most of 
the possible solutions to the energy and momentum conservation 
equations \rff{42}
remain the same as those discussed in the previous paragraph, i.e. in principle
one can have `collinear exchange' and `no collision'
motion with an arbitrary number of wormhole traversals, or `velocity
exchange' motion with a single wormhole traversal, depending on the
direction of $\vi\dot = c\vec{k}_i$ ($\v\vec{k}_i\v=1$).
The only slight difference is with the case of `mirror exchange'
(i.e., $\vi$ directed away from both wormhole mouths A and B), whose only
non trivial solution is given by Eq. \rff{50} replaced by 
\begin{equation}
\left. \begin{array}{l} 
(v_1^{~\prime})_x=(v_{n+1})_x\simeq finite <c,
\\[0.5em] 
(v_{n+1}^{~\prime})_x=(v_i)_x\simeq c[1-(\epsilon_2)^2/2], 
\\[0.5em] 
(v_1^{~\prime})_y=-(v_{n+1})_y\simeq finite <c,
\\[0.5em] 
(v_{n+1}^{~\prime})_y=-(v_i)_y\simeq \epsilon_1 c,
\\[0.5em]
\g_i(v_i)_y=-\g_{n+1}(v_{n+1})_y \equiv \epsilon =finite,
       \end{array}
\right.  
\label{82b}
\end{equation}
with $\v\epsilon_1\v<\v\epsilon_2\v\ll 1$ and $\v\epsilon_2\v/\v\epsilon_1\v
\simeq (1+c^2/\epsilon^2)^{1/2}$.

\paragraph{c) Cauchy data with ultrarelativistic velocity II: 
$\gamma_k\rightarrow \infty\;\; ,\;\; k=i, ... n+1$}
\hfill\break

The case in which both the initial velocity and that of all other copies of 
the particle are ultrarelativistic 
is again similar to the case of initial Cauchy data with generic
relativistic velocity, but only `collinear exchange', `no collision'
(eventually with multiple wormhole traversals) and `velocity exchange'
(only with one wormhole traversal) motions are in principle
possible, while the `mirror exchange' solutions are no longer possible.
This is easily seen from the fact that, for the case in which all $\gamma_k
\rightarrow \infty$ ($k=i, ... n+1$), introducing for the velocities
the notation
\be
\left. \begin{array}{lcl} 
\vec{v}_k\equiv c\vec{k}_k &\;\; ; \;\;  \v\vec{k}_k\v=1 &\;\; ; \;\; 
k=1, ... n+1,
       \end{array}
\right.  
\lb{86}
\ee
the conservation Eqs. \rff{42} reduce to the single condition
\be
\vec{k}_i+\vec{k}_{n+1}=\vec{k}_{1}^{~\prime}+\vec{k}_{n+1}^{~\prime},
\lb{87}
\ee
which only admits solutions with `velocity' and `collinear' exchange 
`topologies',
but not `mirror' type ones.

\section{The Cauchy initial value problem}

We turn now to the {\it global}
 analysis of the stationarity problem for the action 
\rff{6}, representing the motion of a relativistic particle passing
 through a wormhole `time machine' an arbitrary number of times,
self-colliding once and subject
to a given set of fixed initial Cauchy data.

In the previous Section we discussed the possible solutions
to the {\it local} energy and momentum conservation laws which can
be deduced by imposing the stationarity of the action \rff{6}.
Our task is now to verify whether there exist globally self-consistent
motions which are stationary points for
the action \rff{6} for a generic set of Cauchy initial data, in particular for
any given modulus and direction of the initial velocity (and for any given
initial position $\ri$).
In doing this we will  be proving that, 
for the model of a relativistic particle self-interacting 
via the `hard-sphere' potential \rff{2} and constrained to traverse a
wormhole `time machine', the whole set of classical trajectories
which are globally self-consistent are those which can be derived by simply
imposing the `Principle of stationary action'.
In this sense, we would be thus extending to the case 
of relativistic motion of the particle the result,
previously obtained in Refs. TM-(I,II) for the case of the non-relativistic
motion, that the `Principle of 
self-consistency' is a consequence of the more general `Principle of
stationary action'.

Finally, we will also study the multiplicity 
(defined as the number of  self-consistent classical trajectories 
beginning with a fixed set of initial Cauchy data, see Ref. \cite{ekt})
of these stationary points,
and compare our results with those obtained in the 
non-relativistic model of Ref. \cite{ekt}.

Our discussion of the Cauchy problem will be performed by considering
separately the possible values of the modulus of the initial
velocity of the relativistic particle (i.e., `generic relativistic', when
$\gamma_i \sim finite$, `ultrarelativistic (I)', when $\gamma_i \rightarrow 
\infty$ but $\gamma_k\sim finite $ for $k\not = i$, and
`ultrarelativistic (II)', when all $\gamma_k \rightarrow \infty$ for $k=i, ... 
n+1$) and, for each of these cases, the various possible directions of the 
same velocity (i.e., particle initially pointing away from both wormhole 
mouths, particle pointing towards mouth B and particle pointing towards 
mouth A).
In the following subsections we summarize the main results of such an
analysis, while for a more detailed discussion of the trajectories
for each of the fixed initial Cauchy data we refer to Appendix A.2.

\subsection{Cauchy data with generic relativistic velocity:\hfill\break
$\gamma_i\sim finite$}
\paragraph{a) $\vi$ pointing away from mouths A and B}
\hfill\break
If the initial velocity of the particle is smaller than $c$
and pointing away from
both wormhole mouths, a simple analysis of the various
solutions of the energy and momentum conservation equations
\rff{42} shows that the only possible globally self-consistent 
`topologies' for the particle motion are those 
corresponding to the
self-collision of the particle in Region II with  `mirror exchange'
or `velocity exchange' of velocities 
(plus the `trivial' motion in which the 
particle does not enter the wormhole, with multiplicity one).
As it is shown in Appendix A.2, for these kinds of initial Cauchy data
a unique (non trivial) stationary point for the action, and therefore a 
unique classical motion (Eqs. \rff{b7b} and \rff{b17}-\rff{b18} for the `mirror
exchange' case, and Eqs. \rff{ve3}-\rff{ve4} for the `velocity exchange'
case), is possible for 
any fixed number $n$ (only $n=1$ for
the `velocity exchange' case) of wormhole traversals
by the relativistic particle, subject to the constraints \rff{b8}, 
\rff{b20c}-\rff{b21} and \rff{me5a}-\rff{me5aa} 
(for the `mirror exchange' case,
in the `gauge' fixed by Eqs. \rff{b5} and \rff{b8b}),
or \rff{ve5} and \rff{ve7}-\rff{ve8} (for the `velocity exchange' case,
in the `gauge' fixed by Eqs. \rff{ve2}).

Since both `mirror exchange', `velocity exchange' and `no entrance'
motions are possible in this case, and the number of possible
wormhole traversals is in principle arbitrary, the multiplicity
of the trajectories for this case is at least one
(in general finite, as discussed in details 
in Appendix A.2), and it becomes infinite (again, 
in the frame fixed by Eqs. \rff{b5} and \rff{b8b})
provided that, e.g., the constraints \rff{me4}-\rff{me5} and \rff{b25} are
satisfied (see the discussion in Appendix A.2).
Thus, generally, the Cauchy initial value 
problem is classically ill-posed (far too many solutions).

\paragraph{b) $\vi$ pointing towards mouth B}
\hfill\break
When the relativistic particle is initially moving towards wormhole
mouth B, the only possible, globally self-consistent, `topology'
for the particle motion  is that corresponding to the case
of `no collision'.
Also in this case it can be shown (see Appendix A.2 for details) that 
there is a unique stationary point for the action, and therefore a 
unique classical trajectory (Eqs. \rff{70b}-\rff{69}),  
for any fixed number $n$ of wormhole traversals
by the relativistic particle (provided condition \rff{72} holds).

Since the number of wormhole traversals is arbitrary,
the multiplicity of the trajectory is in general
finite (at least one), and it becomes infinite if, e.g., condition \rff{75a}
is satisfied.
Again, generally the Cauchy initial value 
problem is classically ill-posed.

We would like to stress that, among the `no collision' cases, we are also
including trajectories which are treated like that only because of our 
approximation that the wormhole mouths are pointlike.
Indeed, if the Cauchy initial data are specified in such a way that the
original and the final copies come to the same space point simultaneously,
then there might be a collision, but only a `glancing' one.
After such a `gentle' collision the original copy would move along a
slightly altered trajectory which still takes it into mouth B because of
the finite size of this mouth.
In other words, this case would actually correspond to the class of the
so called `dangerous trajectories' described in Ref. \cite{ekt}.
Only when we take the limit of pointlike mouths we should treat this case
among the `no collision' ones.

\paragraph{c) $\vi$ pointing towards mouth A}
\hfill\break
Finally, when the relativistic particle is initially heading towards
wormhole mouth A, only `collinear exchange' or `no collision' motions
are possible.\footnote{In particular, the `no collision' case distinguishes
itself from all other motions described in this paper since the particle
travels {\it forward} in time while traversing the wormhole.}
In this case, the trajectory for which the action \rff{6} is stationary
is also unique (Eqs. \rff{c12}-\rff{c10}, in the `gauge' \rff{c4}, for
the `collinear exchange' motion, and Eqs. \rff{70b} - with subscripts A and B
exchanged - and \rff{k3} for the `no collision' motion), 
for each fixed number $n$ of 
wormhole traversals (provided, for the `collinear exchange' case, 
the constraints \rff{c13}-\rff{c13a} of Appendix A.2 are satisfied).

Since the `no collision' motion is always possible for arbitrary $n$
(only provided $\v\vi\v <c$), in this case
the multiplicity of solutions to the equations of motion is 
always infinite.
The Cauchy initial value problem is always classically ill-posed.

We should note here that for the `no collision' trajectories 
the multiplicity of solutions is in some sense fictitious and an
artifact of our approximation of treating the wormhole mouths as pointlike.
In particular, if we take into account the finite size of the mouths
and we fix the wormhole internal structure such as, e.g., the traversal
rules of Ref. \cite{ekt} are satisfied, then for different $n$'s one
would need slightly different Cauchy initial data.
On the other hand, the finite size of the mouths would remove the degeneracy
of the `no collision' trajectories, since also the motions with `mirror
exchange' collisions would become possible.
 
\subsection{Cauchy data with ultrarelativistic velocity I:\hfill\break 
$\gamma_i\rightarrow \infty ~~;~~\gamma_k\sim finite ~~,~~ k\not = i$}

The discussion about the cases of a particle traversing the wormhole
`time machine' and whose initial velocity is ultrarelativistic 
(but the velocities of all other copies of the particle are only relativistic,
i.e. with $\gamma_k\sim finite$ for $k\not = i$) is 
essentially the same as that made in the previous Section, with
the only caveat that everywhere in the formulas for the trajectories
one should, of course, substitute $\vi$ by $c\vec{k}_i$ ($\v\vec{k}_i\v=1$)
and, for the `mirror
exchange' type of collision, use the `gauge' fixing condition
\rff{b33} of Appendix A.2 instead of Eq. \rff{b8b}.
For any given Cauchy data, any given `topology' of the motion
(`no collision', `mirror exchange', `velocity exchange'
and `collinear exchange') and any fixed number of wormhole traversals,
the multiplicity of the globally self-consistent
classical trajectory for which the
action is stationary is one, 
and therefore generically finite when allowing for an arbitrary number
of such traversals (provided certain constraints are satisfied).
For case {\bf a)}, i.e. an initial velocity pointing away from both wormhole
mouths, the multiplicity of classical trajectories is at least one and
becomes infinite if 
the constraints \rff{me5a}-\rff{me5aa} and, e.g., \rff{b25} are satisfied.
For case {\bf b)}, i.e. an initial velocity pointing towards mouth B, 
 condition \rff{75a} reduces to the first of
conditions \rff{duea}
for the existence of the `time machine'.
Therefore, the multiplicity of the classical trajectories 
is always infinite in this case.
In case {\bf c)} the multiplicity is also always infinite.
For any choice of the initial data (except when both any of the constraints
\rff{b8}, \rff{b20c}-\rff{b21}, \rff{me5} - in the `gauge' \rff{b5}, 
\rff{b8b} -
and any of the constraints \rff{ve5}, \rff{ve7}-\rff{ve8} - in the `gauge'
\rff{ve2} - are violated, in which case the multiplicity of trajectories
is one and
the Cauchy problem well defined) the Cauchy initial value 
problem is classically ill-posed.

\subsection{Cauchy data with ultrarelativistic velocity II:\hfill\break
$\gamma_k\rightarrow \infty\; ,\; k=i, ... n+1$}

The only change with respect to the discussion in the previous Section
for the case in which the velocities 
of all the copies of the particle (including the initial velocity)
are ultrarelativistic,
is in the impossibility of having `mirror exchange' type of solutions
to the equations of motion (see Eq. \rff{87} in Section 3.3.1).
Therefore, for the case of Cauchy data in which the particle is initially
pointing away from both wormhole mouths, the only possible trajectories
are the trivial one where the particle never enters the wormhole 
(multiplicity one) and 
that corresponding to `velocity exchange' (multiplicity one if conditions
\rff{ve5} and \rff{ve7}-\rff{ve8} hold).
For all choices of the initial data (except those violating any of the
constraints \rff{ve5}, \rff{ve7}-\rff{ve8}, for which the multiplicity of 
trajectories is one  and
the Cauchy problem well defined)
 the Cauchy initial value 
problem remains classically ill-defined.
 
We have thus proved that,
for the model of a particle which is constrained to traverse (albeit
an arbitrary number of times) a wormhole
`time machine' geometry, to self-interact via the
`hard-sphere' potential \rff{2} and subject to any of the initial
Cauchy data described in Sections 4.1-4.3,  the whole set of
classical trajectories which are
globally self-consistent can be directly and simply recovered
by imposing the `Principle of stationary action'.
However, the multiplicity of trajectories starting with fixed initial data 
is generically finite, if not even infinite, thus making the Cauchy initial
value problem classically ill-posed.

\section{Discussion}
\label{Sect:Discussion}

The analysis of the Cauchy initial value 
problem for some simple physical systems, e.g. represented
by free classical fields or self-interacting particles, evolving in the
background of a spacetime containing CTCs has been recently tackled by the 
authors of Refs. [9-11].
\footnote{
For the description of the classical and quantum initial value problems
for chronology violating spacetimes in which space consists of a finite
number of space points, see also Ref. \cite{few}.}
In particular, the authors of Ref. \cite{con} considered the Cauchy 
initial value problem
for the evolution of a classical, massless scalar field $\phi$ in the
presence of CTCs, showing that in general the `Principle of self-consistency'
constrains those initial data for the field $\phi$ which are posed 
in the future of the Cauchy horizon where the CTCs reside.
These constraints appear to be mild in the sense that the initial data
can be fixed arbitrarily in some neighbourhood of any event, while
being adjusted elsewhere in order to guarantee a globally self-consistent
evolution.
The authors of Ref. \cite{ekt}, instead, mainly focused their attention
to the Cauchy initial value 
problem for the motion of a non-relativistic, classical
`billiard ball'-like particle in the background of a wormhole
`time machine'.
Allowing for an arbitrary number of wormhole traversals by the particle,
it was claimed that almost all initial trajectories 
 have infinite multiplicity and
make the Cauchy initial value problem ill-posed (far too many solutions).
In particular, for a fixed initial path and speed (i.e., fixed $\v\vi\v$,
$\psi_A$ and $h$, the last two quantities respectively representing
the angle between $\vi$ and the mouths line of centers, and the initial
impact parameter along $\vi$ with respect to mouth B) of the particle, coplanar
motion with respect to the wormhole mouths, and when the constraints
\footnote{Since $L$ is the distance, along the $\vi$ direction, from 
the collision to the point of closest approach to mouth B, this means
that the collision occurs {\it far} from the wormhole.} 
\bea
&&\v\vi\v >\v\rb-\ra\v/\t,  
\non \\
&&L\gg \v\rb-\ra\v; ~~~L\gg h
\lb{ektm}
\eea
are satisfied,
it was shown that the number of solutions for a single `mirror exchange'
collision is infinite. 
Only the trajectory for which the particle is initially at rest
far from the wormhole appears to have multiplicity one, while no
evidence was found for trajectories with zero multiplicity.
\footnote{The issue of the possible existence of zero multiplicity,
i.e. globally self-inconsistent, classical trajectories in the 
background of a wormhole `time machine' has been also addressed in
Refs. \cite{rama}.}
The analysis was also generalized to the case of slightly non coplanar
motion, confirming that infinite multiplicity of solutions should be
generic.

Similarly, in this paper we considered the Cauchy initial value 
problem for the motion of
a pointlike relativistic particle which can traverse a wormhole `time machine'
several times and self-interacts via a short range, `hard-sphere' potential.
We showed that all of the trajectories of the 
relativistic
particle which are globally self-consistent can be found by simply
imposing the `Principle of stationary action', thus confirming
our earlier conjecture (formulated in the context of a fixed boundary
data problem for a non-relativistic motion, Refs. TM-(I,II)) that the
`Principle of self-consistency' is a natural consequence of more
fundamental physical principles.
In particular, we also addressed the issue of the classical 
multiplicity of such
trajectories, finding that it is always at least one, in general finite and, 
if certain constraints are satisfied by the initial
Cauchy data, even infinite.
The set of constraints leading to infinite multiplicity of solutions
is consistent with that previously found in Ref. \cite{ekt} for the
case of non-relativistic motion.
However, as explained in more details in Appendix A.2, the results
of our analysis are apparently in contrast with the claim of Ref. 
\cite{ekt} that infinite multiplicity of trajectories is a generic
property of any set of initial Cauchy data.
The existence of a non empty set, among our solutions, of trajectories
having finite multiplicity can be shown to be the direct consequence
of the different definition of the initial data for the Cauchy problem
(the authors of Ref. \cite{ekt} only fix the initial path and leave
one of the components of $\ri$ as an arbitrary parameter of the motion)
and, moreover, of the kinematical constraints on the velocities of
the particle ($\v\vec{v}_k\v<c$, $k=i, n+1$, for a particle initially pointing
away from both wormhole mouths A and B) or of the conditions for the existence
of the `time machine' (for a particle initially moving towards mouth B).
These constraints  apparently were not considered
in Ref. \cite{ekt}.

On the other hand, our conclusion that the Cauchy initial value problem
for the motion of a relativistic `billiard ball'-like particle in the spacetime
containing CTCs is generally
ill-posed (too many solutions) extends the
non-relativistic results of Ref. \cite{ekt}.

Even if in our analysis, which is purely classical, we have considered for
simplicity only those motions entailing at most a single self-collision,
we expect that multiple-collision solutions should further increase
the multiplicity of trajectories, in agreement with what suggested
in Ref. \cite{ekt}.

Finally, in Refs. \cite{ekt} and \cite{last} it was claimed that,
when taking into account the effects of quantum mechanics, the classically,
non-relativistic ill-posed Cauchy problem should become quantum mechanically
well-posed, in the sense that the probability distributions for the
outcomes of measurements (predicted in the context of a path integral,
sum-over-histories formulation, see Ref. \cite{hartle}) turn out to be 
unique.\footnote{Although the quantum evolution of self-interacting states
propagating in spacetimes with CTCs might be non-unitary, see, e.g.,
Refs. \cite{few},\cite{unit} and \cite{mn2}.}
The question whether similar quantum effects can also `beneficially'
apply to the ill-posedness of the Cauchy initial value
problem for the relativistic
model presented here is an interesting and still open issue which we hope
to address in a future publication.

\vspace{33pt}

\noindent {\Large \bf Acknowledgements}{\vspace{11pt}}

This work is supported in part by Danish Natural Science Research
Council through grant N9401635 and in part by Danmarks
Grundforskningsfond through its support for the establishment of the
Theoretical Astrophysics Center.
A.C.'s research was partly funded by TAC and also by a JSPS postdoctoral 
fellowship, under contract No. P95196.

\newpage

\appendix
\section{Appendix}
\subsection{The `hard-sphere' potential}
\label{App:Hard}

In this appendix we prove some interesting properties for the
`hard-sphere' potential of Eq.~\rff{2}.\footnote{These results are valid
for the generic case of motion in three spatial dimensions.}

$\bullet$ {\it Energy conservation}

For the potential \rff{2}, the energy conservation Eq.~\rff{13a} is
apparently ill defined at the collision event.
What, however,  this equation clearly says is that,
in the region $r<r_s$, the motion is not classically allowed, as
the total relativistic energy would be infinitely large and negative.
Noting that the relativistic $\gamma$ factors satisfy the following
properties
\be
\left. \begin{array}{l}
(\g_k\dot{ \vec{r}_k})^{\cdot}=\g_k^3\ddot{\vec{r}_k},
\\[0.5em]
\dot \g_k=\dot{\vec{r}}_k\cdot \ddot{\vec{r}}_k\g_k^3/c^2,
      \end{array}
\right.
\lb{a1}
\ee
and assuming that the classical motion proceeds until $r=r_s$ (i.e., the
collision takes place at time $\to$), we can now integrate the left hand sides
of Eqs.~\rff{9} around $r_s$, obtaining
\bea
m\int_{r_k(\to-\d t_1)}^{r_k(\to+\d t_2)}d\vec{r}_k~\cdot
(\g_k\dot{\vec{r}}_k)^{\cdot}
&=&m\int_{\to-\d t_1}^{\to+\d t_2}dt~\dot{\vec{r}}_k\cdot
\ddot{\vec{r}}_k\g_k^3=mc^2\int_{\to-\d t_1}^{\to+\d t_2}dt~\dot\g_k
\non \\
&=&mc^2[\g_k(\to+\d t_2)-\g_k(\to-\d t_1)],
\lb{a2}
\eea
for $k=1, n+1$ and where $\d t_1, \d t_2 >0$.
Summing Eqs. \rff{a2} for $k=1$ and $k=n+1$ we then get
\bea
I&\dot =&m\left [\int_{r_1(\to-\d t_1)}^{r_1(\to+\d t_2)}d\vec{r}_1~\cdot
(\g_1\dot{\vec{r}}_1)^{\cdot}+
\int_{r_{n+1}(\to-\d t_1)}^{r_{n+1}(\to+\d t_2)}d\vec{r}_{n+1}~\cdot
(\g_{n+1}\dot{\vec{r}}_{n+1})^{\cdot}\right ]
\non \\
&=&mc^2\{[\g_1(\to+\d t_2)+\g_{n+1}(\to+\d t_2)]
-[\g_1(\to-\d t_1)+\g_{n+1}(\to-\d t_1)]\}.
\lb{a2a}
\eea
Moreover, when integrating on the right hand sides of Eqs. \rff{9},
one obtains
\bea
I&=&\hat V\left 
[\int_{r_{n+1}(\to-\d t_1)}^{r_{n+1}(\to+\d t_2)}d\vec{r}_{n+1}~\cdot
{\r\over r}\d (r-r_s)-\int_{r_1(\to-\d t_1)}^{r_1(\to+\d t_2)}d\vec{r}_1~\cdot
{\r\over r}\d (r-r_s)\right ]
\non \\
&=&\hat V\int_{r_s+\epsilon_1}^{r_s+\epsilon_2}dr~\d (r-r_s)=0
\lb{a3}
\eea
(with $\epsilon_1, \epsilon_2>0$).
Comparing Eqs. \rff{a2a} and \rff{a3} we finally get
the conservation law for the relativistic energy before and after the
collision
\be
m(\g_1+\g_{n+1})c^2=const,
\lb{a4}
\ee
as anticipated.
\vfill\eject
$\bullet$ {\it Zero contribution to classical action}

Let us consider the contribution of the potential \rff{2} to the action
\rff{6} evaluated along the classical trajectories \rff{387} or
\rff{314}.

In the `no collision' case, denoting with $t_m$ and $r_m$ the time
and position of minimum distance
between the two copies of the particle, we have, for the potential
\rff{2},
\begin{eqnarray}
\int_{\tbn}^{\tbuno+\t}dt~V(r)\biggr\v_{cl} &=&
\hat V\left [\int_{\tbn}^{t_m}dt~\theta(r_s-r)
+\int_{t_m}^{\tbuno+\t}dt~\theta(r_s-r)\right ]
\non \\
&=& \hat V\left [\int_{r(\tbn)}^{r_m}{dr\over \dot r}~\theta(r_s-r)
+\int_{r_m}^{r(\tbuno+\t)}{dr\over \dot r}~\theta(r_s-r)\right ] =0,
\lb{005}
\end{eqnarray}
since obviously, in the `no collision' case, $r_m, r(\tbn)$ and $r(\tbuno+\t)$ 
are
always greater than $r_s$.

The proof proceeds along similar lines for the collision case (see, i.e.,
Ref. TM-I).

In conclusion, the contribution of the `hard-sphere' potential $V$
to the classical action can always be neglected both in the collision
and `no collision' cases.

\subsection{Trajectories for the coplanar motion}
\label{App:Traj}

\paragraph{a) $\vi$ pointing away from mouths A and B}
\hfill\break
\vskip 0.3cm
\paragraph{i) Generic relativistic velocity: $\gamma_i \sim finite$}
\hfill\break
$\bullet$ {\it `Mirror exchange of velocities'}

When the Cauchy data for the relativistic particle are such that the
initial velocity is pointing away from both wormhole mouths, with $\gamma_i
\sim finite$, one possible local 
solution of the conservation Eqs. \rff{42} is given 
by the `mirror exchange' type of collision.
In this case, assuming $n$ traversals of the particle through the wormhole 
`time
machine', and in the frame specified by Eq. \rff{48} of Section 3.3.1, we can
write down in components the trajectories for the two copies of the 
particle in Region II as  
\be
\left. \begin{array}{l}
x_1(t)=(v_i)_xt+(b_1)_x,
\\[0.5em]
y_1(t)=(v_i)_yt+(b_1)_y,
\\[0.5em]
x_{n+1}(t)=(v_{n+1})_xt+(b_{n+1})_x,
\\[0.5em]
y_{n+1}(t)=-\g_i(v_i)_yt/\g_{n+1}+(b_{n+1})_y,
\\[0.5em]
x_1^{~\prime}(t)=(v_{n+1})_xt+(b_1^{~\prime})_x,
\\[0.5em]
y_1^{~\prime}(t)=\g_i(v_i)_yt/\g_{n+1}+(b_1^{~\prime})_y,
\\[0.5em]
x_{n+1}^{~\prime}(t)=(v_i)_xt+(b_{n+1}^{~\prime})_x,
\\[0.5em]
y_{n+1}^{~\prime}(t)=-(v_i)_yt+(b_{n+1}^{~\prime})_y.
      \end{array}
\right.
\lb{b4}
\ee
We then further fix our coplanar coordinate system by choosing the collision
point $\ro$ as its origin, i.e. we take
\be
\xo=\yo=0,
\lb{b5}
\ee
and also define, for simplicity of notation, the following quantity
\be
\rho\equiv (n-1){\v \rb-\ra\v\over \v\vnpuno\v}.
\lb{b6}
\ee
Since the `unknowns' in the problem \rff{b4} are twelve 
($(b_{1})_{x, y}$, $(b_{1}^{\prime})_{x, y}$, $(b_{n+1})_{x, y}$, 
$(b_{n+1}^{\prime})_{x, y}$, $(v_{n+1})_{x, y}$, 
$\to$ and $\tbuno$), while the conditions
to be imposed coming from the Cauchy and boundary data (Eqs. \rff{unoa},
\rff{unob} and \rff{311b}) are fourteen, the solutions will be constrained.
In particular, it is quite straightforward to formally solve for the 
trajectory parameters as
\be
\left. \begin{array}{l}
(b_1)_x=(b_{n+1}^{~\prime})_x=\xi-(v_i)_x\ti,
\\[0.5em]
(b_1)_y=-(b_{n+1}^{~\prime})_y=\yi-(v_i)_y\ti,
\\[0.5em]
(b_{n+1})_x=(b_1^{~\prime})_x=(\xa-\xb)[\ti-\xi/(v_i)_x]/(n\t-\rho),
\\[0.5em]
(b_{n+1})_y=-(b_1^{~\prime})_y=(\ya+\yb)[\ti-\xi/(v_i)_x]/(n\t-\rho),
\\[0.5em]
(v_{n+1}^{~\prime})_x=(v_i)_x,
\\[0.5em]
(v_{n+1}^{~\prime})_y=-(v_i)_y,
\\[0.5em]
(v_{n+1})_x=(v_1^{~\prime})_x=(\xb-\xa)/(n\t-\rho),
\\[0.5em]
(v_{n+1})_y=-(v_1^{~\prime})_y=-(\ya+\yb)/(n\t-\rho),
      \end{array}
\right.
\lb{b7a}
\ee
with the times of collision and of first entrance
of the particle into wormhole mouth B 
respectively given by
\bea
\to&=&\ti-\xi/(v_i)_x,
\non \\
\tbuno&=&\to-\t+{\xb\over (\xb-\xa)}(n\t-\rho),
\lb{b7b}
\eea
while the constraints can be written as
\be
\left. \begin{array}{l}
(v_i)_x/(v_i)_y=\xi/\yi,
\\[0.5em]
\yb/\xb=-\ya/\xa,
      \end{array}
\right.
\lb{b8}
\ee
plus the `gauge' condition
\be
(\ya+\yb)\gamma_{n+1}/(n\t-\rho)=
(v_i)_y\gamma_i.
\lb{b8b}
\ee

At this level, the expressions for $([b\div v]_{n+1})_{x, y}$
in Eqs. \rff{b7a}, and for $\tbuno$ in the second of
Eqs. \rff{b7b},
are still formal and implicit, since they depend on the parameter 
$\rho$, which, 
by its definition \rff{b6}, is a function of the velocity $\v\vnpuno\v$.
In order to solve for the components $(v_{n+1})_x$ and 
$(v_{n+1})_y$, we first rewrite the last two of Eqs. \rff{b7a} introducing
the explicit form of $\rho$, Eq. \rff{b6}, and obtaining  
\be
\left. \begin{array}{l}
(v_{n+1})_x(n-1)\v\rb-\ra\v=\v\vnpuno\v[\xa-\xb+n\t(v_{n+1})_x],
\\[0.5em]
(v_{n+1})_y(n-1)\v\rb-\ra\v=\v\vnpuno\v[\ya+\yb+n\t(v_{n+1})_y].
      \end{array}
\right.
\lb{b11}
\ee
Dividing member by member the two equations in \rff{b11}, we have 
\be
\left. \begin{array}{l}
(v_{n+1})_x(n-1)\v\rb-\ra\v=\v\vnpuno\v[\xa-\xb+n\t(v_{n+1})_x],
\\[0.5em]
(v_{n+1})_y=(v_{n+1})_x(\ya+\yb)/(\xa-\xb).
      \end{array}
\right.
\lb{b12}
\ee
The system of equations \rff{b12} can be finally solved for
$(v_{n+1})_{x, y}$ squaring the last of Eqs. \rff{b12}, using the first
of Eqs. \rff{b12} and the second of conditions \rff{b8} in order to
eliminate the dependence on $\v\vnpuno\v$, and finally
making use of the following constraints 
\be
\left. \begin{array}{l}
\xb(n\t-\rho)/(\xb-\xa)>0,
\\[0.5em]
\xa(n\t-\rho)/(\xb-\xa)<0,
      \end{array}
\right.
\lb{b10}
\ee
which are a consequence of the first two of the
time ordering constraints \rff{dueaa} 
applied to the `mirror exchange' solutions \rff{b7b} (moreover,
the last of constraints \rff{dueaa} gives $\xi/(v_i)_x<0$).
The explicit result which is consistent with the conditions \rff{b10}
is given by
\be
\left. \begin{array}{l}
(v_{n+1})_x=[(1-n)\v\rb-\ra\v\xa+(\xb-\xa)r_A]/n\t r_A,
\\[0.5em]
(v_{n+1})_y=\ya(v_{n+1})_x/\xa.
      \end{array}
\right.
\lb{b13}
\ee
In particular, noting from the second of Eqs. \rff{b13} that we have 
\be
\v\vnpuno\v=-(v_{n+1})_x{r_A\over \xa},
\lb{b15}
\ee
we can also explicitly write down for the parameter $\rho$ 
\be
n\t-\rho={n\t(\xb-\xa)r_A\over [(1-n)\v\rb-\ra\v\xa+(\xb-\xa)r_A]},
\lb{b16}
\ee
and, therefore, from Eq. \rff{b7b},
we find the explicit solution for $\tbuno$ as
\be
\tbuno=\to-\t+{n\t\xb r_A\over [(1-n)\v\rb-\ra\v\xa+(\xb-\xa)r_A]}.
\lb{b17}
\ee

Using all these results into Eqs. \rff{b4}, 
the final explicit form of the classical
trajectories for the motion with `mirror exchange' collision can be
written as 
\be
\left. \begin{array}{l}
x_1(t)=x_{n+1}^{~\prime}(t)=(v_i)_x[t-\ti +\xi/(v_i)_x],
\\[0.5em]
y_1(t)=-y_{n+1}^{~\prime}(t)=(v_i)_y x_1(t)/(v_i)_x,
\\[0.5em]
x_{n+1}(t)=x_1^{~\prime}(t)=[(1-n)\v\rb-\ra\v\xa+(\xb-\xa)r_A]x_1(t)
/n\t r_A(v_i)_x,
\\[0.5em]
y_{n+1}(t)=-y_1^{~\prime}(t)=\ya x_{n+1}(t)/\xa.
      \end{array}
\right.
\lb{b18}
\ee
Moreover, since for such trajectories one can easily show that $n\t -\rho>0$,
it is straightforward to check that the possible solution for the system
of constraints \rff{b10} is  given by 
\be
\xa/\xb <0.
\lb{b20c}
\ee

Finally, we have a further condition coming from the kinematical
limit $\v\vnpuno\v< c$
which, when used into Eqs. \rff{b13} and \rff{b15}, leads to the final
constraint
\be
{[(n-1)\v\rb-\ra\v\xa+(\xa-\xb)r_A]\over n\t \xa}< c.
\lb{b21}
\ee
Moreover, from  Eqs. \rff{b13}-\rff{b15} and the constraint \rff{b20c}
it is easy to see that the condition \rff{28} for the existence of the
`time machine' is also trivially satisfied.
 
From our analysis we can conclude that the Cauchy initial value 
problem for the motion of a 
relativistic particle constrained to traverse a wormhole `time machine' 
$n$-times ($n$ fixed), to have an initial velocity pointing away from both
wormhole mouths A and B and self-interacting via a `mirror exchange' 
collision, has a unique, globally self-consistent 
solution, given by the first of Eqs. \rff{b7b} and Eqs. \rff{b17}-\rff{b18},
 provided that the initial Cauchy data are chosen to
satisfy, i.e., the first of conditions \rff{b8} and condition \rff{b21}
holds.
Moreover, the second of conditions \rff{b8} and condition \rff{b8b}
can be interpreted as
fixing the actual coordinates of the collision point in an arbitrary
frame. 
This is because such conditions are not written in an
invariant form under a global change of coordinates in the 2-d spatial
frame where the wormhole mouths are at rest, so that changing the
values of, e.g., $\xa$ and $\ya$ (leaving the other coordinates fixed)
in \rff{b8} and \rff{b8b} is actually equivalent to change the 
position of $\ro$ relative to the wormhole mouths A and B.
The trajectories are further restricted to satisfy the 
constraint \rff{b20c}.

The multiplicity of the trajectories can be seen to depend
on conditions \rff{b8b} and \rff{b21}, which 
involve the initial velocity $\vi$ (condition \rff{b21} implicitly
depends on $\vi$ via Eq. \rff{b8b}).
In particular, it is instructive to rewrite the constraint \rff{b8b},
using Eqs. \rff{b8}, \rff{b13} and \rff{b16}, extracting the initial velocity
$\v\vi\v$ as a function of $n$ and the other trajectory parameters, i.e.
\be
\v\vi\v^2=c^2[\alpha f(n)/(1+\alpha f(n))],
\lb{me1}
\ee
where we have defined the function
\be
f(n)~\dot = ~[(n-1)a+b]^2/\{n^2d-[(n-1)a+b]^2\}
\lb{me2}
\ee
and the quantities
\be
\left. \begin{array}{l}
\alpha ~\dot =~\ya^2r_i^2/\yi^2 r_A^2~>0,
\\[0.5em]
a~\dot =~\v\rb-\ra\v~>0,
\\[0.5em]
b~\dot=~(1-\xb/\xa)r_A~>0,
\\[0.5em]
d~\dot=~c^2\t^2~>0.
      \end{array}
\right.
\lb{me3}
\ee
It is easy to check from Eq. \rff{me1}
that the velocity $\v\vi\v$ is a monotonic,
decreasing function of the number $n$ of wormhole traversals, with
extrema $[\v\vi\v_{min}]_{n\rightarrow \infty}
=c\{\alpha a^2/[d+(\alpha -1)a^2]\}^{1/2}$,
~$[\v\vi\v_{max}]_{n=1}=c\{\alpha b^2/[d+(\alpha -1)b^2]\}^{1/2}$.

In other words, {\it only} for those initial Cauchy data having a velocity
\be
Min [c, c~g(\xa, \xb-\xa)]>\v\vi\v>c~g(r_A, \rb-\ra),
\lb{me5a}
\ee
with
\be
g(\mu, \nu)\dot ={\v \ya\v\v \nu\v r_i\over \sqrt{c^2\t^2\yi^2\mu^2+
(\xi^2\ya^2-\xa^2\yi^2)\nu^2}}.
\lb{me5aa}
\ee
the constraint \rff{b8b} is solvable, therefore fixing the coordinates of
$\ro$ relatively to A and B.
For instance, if the condition
\be
\v\xi\ya\v \leq \v\yi\xa\v 
\lb{me4}
\ee
holds, then the lower bound on $\v\vi\v$ simplifies as
\be
\v\vi\v>\v\rb-\ra\v/\t.
\lb{me5}
\ee

Finally, let us consider the constraint \rff{b21}, starting 
from the case in which $\xa>0$.
In this case, the constraint \rff{b21} can be rewritten as
\be
n[\v\rb-\ra\v-c\t]\xa<(\xb-\xa)r_A +\v\rb-\ra\v\xa.
\lb{b24a}
\ee
Then, since for the wormhole to act as a `time machine' the first of
conditions \rff{duea} must
hold, it is easy to show that Eq. \rff{b24a} can be satisfied independently
of the number $n$ of wormhole traversals provided that the following 
sufficient condition is satisfied 
\be
\left (1- {\xb\over \xa}\right ){r_A\over c\t}<1,
\lb{b25}
\ee
which also implies, due to the second of Eqs. \rff{b10}, $r_A<c\t$. 
A similar argument works for the case of $\xa<0$, leading again to condition
\rff{b25}.

In other words, as remarked below Eq. \rff{b21}, 
one can see that a change in $n$ actually corresponds, via Eq.
\rff{b8b}, to a change in the position of one of the two coordinates
of the collision point $\ro$ relative to the wormhole mouths A and B.
However, only for those initial velocities satisfying
the constraint \rff{me5a} (and constraint \rff{b21}) 
there is a possible solution to the Cauchy initial value 
problem, each with a different location
of the collision point.

Since, for a given $\vi$, the constraint \rff{b21} can be in general
satisfied only up to a finite number of wormhole traversals, this
means that there is only a finite number of possible locations
for the collision point, and that the multiplicity of the solutions
is in general finite.
On the other hand, if, e.g., the constraint \rff{b25} is also satisfied
by the initial velocity, the kinematical constraint \rff{b21}
is always satisfied (independently of $n$)
and a change in $n$ (for any $n\geq 1$) only
affects the constraint \rff{b8b}, which causes a change in the coordinates
of $\ro$ relative to the mouths A and B.
In particular, when $n\rightarrow \infty$ and assuming $\v\vi\v$ fixed,
one can also 
read Eq. \rff{me1} as giving, e.g., $\ya$ as a function of $\v\vi\v$
(this $\ya$ is finite, as in general also $\xa, \xi$ and $\yi$ are,
see Eqs. \rff{b8}, and there are no further constraints on the initial Cauchy
 data).
Then, only provided that, e.g., the constraints \rff{me4},
\rff{me5} and \rff{b25} are satisfied, the 
multiplicity of trajectories becomes infinite.
We finally note that Eq. \rff{me5} is the same condition assumed
by the authors of Ref. \cite{ekt} (their Eq. (3.4)) to prove that
non-relativistic trajectories with a single `mirror exchange'
self-collision have infinite multiplicity.
However, we have here extended the analysis of Ref. \cite{ekt}
by showing that the stricter condition \rff{me5a} must be satisfied
by the initial data in order that the motion is globally self-consistent.
Moreover, we have shown that initial Cauchy data not satisfying either
condition \rff{me5a} or \rff{b21} lead to a motion with
zero or finite multiplicity, apparently contradicting the claim of
Ref. \cite{ekt} about infinite multiplicity of trajectories being
generic.
The origin of this `mismatch' can be easily traced back partly to the
different choice of initial Cauchy data made by the authors of Ref. \cite{ekt}
(fixed $\vi$ and impact parameter) for describing the particle motion,
and partly to the kinematical constraint \rff{b21}, which was simply 
not considered in Ref. \cite{ekt}.
It can be shown, in fact, that the choice of data made in Ref. \cite{ekt}
(which, in our framework, is essentially equivalent 
to fix $\vi$ and only one component of $\ri$) leads to the same set
of trajectories and constraints as our Eqs. \rff{b17}-\rff{b18}, \rff{b21} 
and \rff{me5a}, with the only difference that now the value of $\to$ is
undetermined and, consequently, the trajectories themselves are implicit
functions of $\to$ (for instance, $x_1(t)=(v_i)_x(t-\to)$).
The point is that with this choice of initial Cauchy data, 
provided the kinematical constraint \rff{b21}
and the gauge condition \rff{b8b} are satisfied {\it at least} for
$n=1$, since one of the components of $\ri$ can be still chosen arbitrarily
(it is a free parameter of the motion), the multiplicity of the trajectories
always turns out to be infinite.


$\bullet$ {\it `Velocity exchange'}

The other possible nontrivial 
motion when the initial velocity of the particle is
heading away from both mouths A and B of the wormhole is that for
which the particle self-collides with `velocity exchange' in Region II
and traverses the wormhole only once (see the discussion in Section 3.3.1).

Taking into account the solution \rff{52} for the conservation equations
at the collision point, the trajectories can be parametrized as
\bea
\runo(t)&=& \vi t +\vec{b}_1,
\non \\
\runop(t)&=& \vdue t +\vec{b}_1^{\prime},
\non \\
\rdue(t)&=& \vdue t +\vec{b}_2,
\non \\
\rduep(t)&=& \vi t +\vec{b}_2^{\prime}.
\lb{ve1}
\eea

Choosing to work in the coordinate frame where 
\be
\left. \begin{array}{l}
\ro = \vec{0},
\\[0.5em]
\ya=\yb=0,
      \end{array}
\right.
\lb{ve2}
\ee
and imposing on the trajectories \rff{ve1} the Cauchy and boundary data 
\rff{unoa} and \rff{unob}, after some simple algebraic steps we finally find 
the following result
\be
\left. \begin{array}{ll}
\runo(t)=\rduep(t)=&\vi(t-\ti)+\ri,
\\[0.5em]
x_2(t)=x_1^{\prime}(t)=&(\xb-\xa)[t-\ti +\xi/(v_i)_x]/\t,
\\[0.5em]
y_2(t)=y_1^{\prime}(t)=&0,
      \end{array}
\right.
\lb{ve3}
\ee
with
\be
\left. \begin{array}{l}
\to=\ti-\xi/(v_i)_x, 
\\[0.5em]
\tbuno=\to+\xa\t/(\xb-\xa),
      \end{array}
\right.
\lb{ve4}
\ee
and subject to the following constraint on the initial Cauchy data
\be
\xi/\yi=(v_i)_x/(v_i)_y.
\lb{ve5}
\ee
Finally, there are the time ordering conditions coming from Eqs. 
\rff{dueaa}
which, for the case of `velocity exchange' collision, read  as
\be
\left. \begin{array}{l}
\xb/(\xb-\xa)>0,
\\[0.5em]
\xa/(\xb-\xa)<0,
\\[0.5em]
\xi/(v_i)_x<0,
      \end{array}
\right.
\lb{ve6}
\ee
and which can be solved if the following constraints hold
\be
\xa/\xb<0
\lb{ve7}
\ee
and
\be
\xi/(v_i)_x<0.
\lb{ve8}
\ee
There are no further constraints coming from the kinematical conditions
$\v\vi\v, \v\vdue\v <c$ (in fact, these are equivalent to the condition
for the existence of the `time machine', i.e. the first of Eqs. \rff{duea}).

Thus, the Cauchy initial value problem for the motion of a 
relativistic particle initially heading away from both 
wormhole mouths A and B, self-interacting via a `velocity exchange' 
collision and traversing the wormhole `time machine' 
only once has a unique, globally self-consistent solution, 
given by Eqs. \rff{ve3} and 
\rff{ve4}, provided that the initial Cauchy data are chosen to
satisfy conditions \rff{ve5} and \rff{ve8}, and the further constraint
\rff{ve7} also holds.
Of course the multiplicity of solutions is one in this case
(zero if any of conditions \rff{ve5}-\rff{ve6} does not hold).
Similarly to the `mirror exchange' case, it can be easily shown that 
the choice of initial data made in Ref. \cite{ekt} would formally
lead to the same set of Eqs. \rff{ve3}-\rff{ve8}, with $\to$ undetermined,
and generally to infinite multiplicity of trajectories.

In conclusion, the Cauchy initial value 
problem for the motion of a relativistic
particle initially heading away from both wormhole mouths A and B
is generally classically ill-posed (i.e. there are far too many
solutions, the trivial one with `no traversal'
of the wormhole, plus one with `velocity exchange' and 
multiple, possibly infinite with `mirror exchange'). 
Only for those initial data not satisfying both any of the constraints
\rff{b8}, \rff{b20c}-\rff{b21}, \rff{me5a} (in the `gauge' \rff{b5}, \rff{b8b})
and any of the constraints \rff{ve5}, \rff{ve7}-\rff{ve8} (in the `gauge'
\rff{ve2}) the trivial motion with `no traversal' is the unique possible 
trajectory and the Cauchy initial value problem is classically well defined.

\paragraph{ii) Ultrarelativistic velocity I: $\gamma_i \rightarrow 
\infty ~~;~~\gamma_k \sim finite ~~,~~k\not = i$}
\hfill\break
In this case the analysis made in the previous paragraph goes through
almost step by step, from the explicit formulas for the times
(Eqs. \rff{b7b} and  \rff{b17} for the `mirror exchange' case, and Eqs.
\rff{ve4} for the `velocity exchange' case) and the trajectories 
(Eqs. \rff{b18} for `mirror exchange' and Eqs. \rff{ve3} for `velocity 
exchange'), 
to the set of constraints (Eqs. \rff{b8}, \rff{b20c}-\rff{b21} and
\rff{me5a}-\rff{b25} 
for `mirror exchange', Eqs. \rff{ve5} and 
\rff{ve7}-\rff{ve8} for `velocity exchange'), albeit the
formal substitution of $\vi$ by $c\vec{k}_i$ ($\v\vec{k}_i\v=1$).
The only difference is that now, for the case of `mirror exchange',
condition \rff{b8b} is
being replaced by
\be
{(\ya+\yb)\over \sqrt{c^2(n\t-\rho)^2-(\xa-\xb)^2}}\simeq
\left \v{\epsilon_1\over \epsilon_2}\right \v sign[(v_i)_y].
\lb{b33}
\ee
Also the results about uniqueness of the (globally self-consistent)
stationary
point at $n$ fixed (for each of the possible `topologies'
of the trajectories), and about multiplicity (generically finite,
infinite when the constraints \rff{me5a} and
\rff{b25} hold) are still valid in the ultrarelativistic case (I).
Generally, the Cauchy initial value problem is classically ill-posed.

\paragraph{iii) Ultrarelativistic velocity II: $\gamma_k \rightarrow 
\infty ~~,~~k=i, ... n+1$}
\hfill\break
Finally, in the 
case in which all the velocities are ultrarelativistic,
only the trajectories with `no traversal' and `velocity exchange'
(with one wormhole traversal)
are possible in general (see the discussion in Section 3.3.1).
In particular, for those initial Cauchy data not obeying any of conditions 
\rff{ve5}-\rff{ve6}, only the trivial `no traversal' motion is possible,
with multiplicity one, and in this case the Cauchy initial value problem
is classically well-posed. 
Otherwise, the multiplicity of solutions is two and 
the Cauchy initial value problem is classically ill-posed.
\vfill\eject
\paragraph{b) $\vi$ pointing towards mouth B}
\hfill\break
\vskip 0.3cm
\paragraph{i) Generic relativistic velocity: $\gamma_i \sim finite$}
\hfill\break
$\bullet$ {\it `No collision'}

The only possible motion when the particle is initially heading towards
wormhole mouth B is that for which the particle enters the
wormhole but experiences `no collision'.\footnote{Regarding the degeneracy of
this case related to our approximations see the end of Section 4.1.b.} 
In this case, the trajectories which are solutions 
of the equations of motion \rff{9} or \rff{42} can be explicitly
parametrized in the
following way
\be
\left. \begin{array}{l} 
\runo(t)=\vi t+\buno \; \; \; ; \; \; \; \ti<t<\tbuno+\t,
\\[0.5em]
\rnpuno(t)=\vnpuno t+\bnpuno \; \; \; ; \; \; \; t>\tbn,
       \end{array}
\right.  
\lb{68}
\ee
where all the velocities are constant and we can write, due to the
conservation Eq. \rff{21},
\be
\vnpuno\equiv \v\vnpuno\v\vec{k}_{n+1}=\v\vi\v\vec{k}_{n+1}\;\; ; \;\;
\v\vec{k}_{n+1}\v=1.
\lb{70a}
\ee
Imposing in Eqs. \rff{68} the initial Cauchy data \rff{unoa}, 
the boundary data \rff{unob} and also using Eq. \rff{27} relating
$\tbn$ to $\tbuno$, we can easily find the explicit form of the trajectories as
\be
\left. \begin{array}{l} 
\runo(t)=\ri+\v\vi\v(\rb-\ri)(t-\ti)/\v\rb-\ri\v,
\\[0.5em]
\rnpuno(t)=\ra+\vec{k}_{n+1}[\v\vi\v(t-\ti+n\t)+(1-n)\v\rb-\ra\v-\v\rb-\ri\v],
       \end{array}
\right.  
\lb{70b}
\ee
and, for the time of first entrance into wormhole mouth B 
\be
\tbuno=\ti-\t+{\v\rb-\ri\v\over \v\vi\v}.
\lb{69}
\ee

The condition for the existence of
the `time machine' (after a fixed - but otherwise arbitrary - number $n$ of 
wormhole traversals), holds provided that the initial Cauchy data
satisfy (see Eq. \rff{28} with $\v\vunop\v=\v\vi\v$)
\be
c>\v\vi\v>(n-1)\v\rb-\ra\v/n\t
\lb{72}
\ee
(for $n=1$ condition \rff{72} is replaced by $\v\vi \v<c$ and the first of 
constraints \rff{duea}).

In conclusion, the Cauchy initial value 
problem for the motion of a relativistic particle
constrained to traverse a wormhole `time machine' $n$-times ($n$ fixed)
and to have an initial velocity pointing towards wormhole mouth B has
a unique, globally self-consistent 
solution provided that the initial Cauchy data are chosen to
satisfy condition \rff{72}.
Moreover, if the initial data are 
also chosen to satisfy the stronger constraint 
\be
\v\vi\v>\v\rb-\ra\v/\t,
\lb{75a}
\ee
then the constraint \rff{72} is easily seen to be satisfied for arbitrary $n$,
thus implying that the multiplicity of the trajectories becomes 
infinite (in this case the kinematical condition $\v\vi\v<c$ does 
not give any further constraint if the first of conditions \rff{duea} holds).
Otherwise, following a reasoning similar to that of the previous paragraph
(i.e., the case of initial Cauchy data with velocity pointing
away from mouths A and B and having a `mirror exchange' collision),
it is possible to show that the multiplicity is always finite (at least 
one if $\v\vi\v<c$ and $\tau>\v\rb-\ra\v/c$).
Again, the condition \rff{75a} for infinite multiplicity of classical
trajectories is the same as that found by the authors of Ref. \cite{ekt},
although, contrarily to their claims, we have shown here that infinite
multiplicity is not generic.
Moreover, fixing the initial data as in Ref. \cite{ekt} would
now imply that the impact parameter is zero, leading formally to the
same trajectories \rff{70b}, but with $\tbuno$ undetermined.
In the ansatz of Ref. \cite{ekt} 
it is easy to see that even just one wormhole
traversal, i.e. when the first of conditions
\rff{duea} for the existence of the `time machine' and the constraint
$\v\vi\v<c$  are satisfied, would be enough to
guarantee infinite multiplicity of solutions.
Generally, the Cauchy initial value problem is classically ill-posed
with both kinds of initial data. 
Only when condition \rff{72} is not satisfied (but still 
$\v\vi\v<c$ and $\t>\v\rb-\ra\v/c$) 
there is a unique possible trajectory (with one wormhole
traversal) and the Cauchy initial value problem is classically well defined.

\paragraph{ii) Ultrarelativistic velocity I/II: 
$\gamma_i \rightarrow \infty ~~;~~ \gamma_k\sim finite ~~,~~ k\not =i ~/~ 
\gamma_k \rightarrow 
\infty ~~,~~k=i, ... n+1$}
\hfill\break
When the initial Cauchy velocity becomes ultrarelativistic 
(and similarly when the velocities of all copies of the particle,
$\vec{v}_k, k=i, ... n+1$, are ultrarelativistic), the
previous results remain formally valid (with, of course,
 $\vec{v}_k$ duly replaced by $c\vec{k}_k, \v\vec{k}_k\v=1$, for 
$k=i, ... n+1$), but, since condition \rff{75a} becomes equivalent to the
condition of existence of the `time machine' (first of Eqs. \rff{duea}),
the multiplicity of the trajectories is always infinite.
For all choices of initial data (with $\v\vi\v<c$ and $\t>\v\rb-\ra\v/c$), 
the Cauchy initial value problem is classically ill-posed.
 
\paragraph{c) $\vi$ pointing towards mouth A}
\hfill\break
\vskip 0.3cm
\paragraph{i) Generic relativistic velocity: $\gamma_i \sim finite$}
\hfill\break
$\bullet$ {\it `Collinear velocities'}

One of the two possible motions in the case of a relativistic particle 
which is initially heading towards wormhole mouth A is that for
which the particle self-collides in Region II under a `collinear exchange'
of velocities, and traverses the wormhole an arbitrary number $n$ of times.
\footnote{Regarding the degeneracy of this case because of our approximations
see the end of Section 4.1.c.}

Using the `collinear exchange' rules \rff{47} and introducing the following 
notation for the velocities of the copies of the
particle in Region II
\be
\left. \begin{array}{llll} 
\vi\equiv \v\vi\v\vec{k}_i&, \vnpuno\equiv -\v\vi\v\vec{k}_i&,
 \vunop\equiv \v\vi\v\vec{k}_1^{\prime}&, \vnpunop\equiv -\v\vi\v
\vec{k}_1^{\prime}
       \end{array}
\right.  
\lb{c2a}
\ee
(where the $\vec{k}_k^{(\prime)}, k=i, 1$ are unit vectors), 
we can write for the
trajectories
\be
\left. \begin{array}{l} 
\runo(t)=\v\vi\v\vec{k}_it+\buno,
\\[0.5em]
\runop(t)=\v\vi\v\vec{k}_1^{~\prime}t+\bunop,
\\[0.5em]
\rnpuno(t)=-\v\vi\v\vec{k}_it+\bnpuno,
\\[0.5em]
\rnpunop(t)=-\v\vi\v\vec{k}_1^{~\prime}t+\bnpunop.
       \end{array}
\right.  
\lb{c3}
\ee
In particular, working in the coordinate frame where
\be
\left. \begin{array}{l} 
\ro=\vec{0},
\\[0.5em]
\ya=\yi=0,
       \end{array}
\right.  
\lb{c4}
\ee
for which we have 
\be
\left. \begin{array}{ll} 
\vec{k}_i=(-[sign(\xi)], 0),&
\\[0.5em]
\vec{k}_1^{~\prime}=(\xb/r_B, \yb/r_B)\;\;\; &;\;\;\;  r_B\equiv \v\rb\v,
       \end{array}
\right.  
\lb{c5}
\ee
one can impose the Cauchy and boundary conditions, Eqs. 
\rff{unoa}
and \rff{unob}, on Eqs. \rff{c3} and finally find for the trajectories
\be
\left. \begin{array}{l} 
x_1(t)=-x_{n+1}(t)=\xi -[sign(\xi)]~\v\vi\v(t-\ti),
\\[0.5em]
y_1(t)=y_{n+1}(t)=0,
\\[0.5em]
x_1^{~\prime}(t)=-x_{n+1}^{~\prime}(t)=-[sign(\xi)]~\xb x_1(t)/r_B,
\\[0.5em]
y_1^{~\prime}(t)=-y_{n+1}^{~\prime}(t)=-[sign(\xi)]~\yb x_1(t)/r_B,
       \end{array}
\right.  
\lb{c12}
\ee
and for the times $\to$ and $\tbuno$
\be
\left. \begin{array}{l} 
\to=\ti+\v\xi\v/\v\vi\v,
\\[0.5em]
\tbuno=\ti-\t+[r_B+\v\xi\v]/\v\vi\v.
       \end{array}
\right.  
\lb{c10}
\ee
It is very easy to check that the solutions \rff{c10} automatically
satisfy the time ordering constraints of Eqs. \rff{dueaa}
(and consequently condition \rff{28} for the existence of the `time
machine').

Moreover, since, as in the case of `mirror exchange' or `velocity exchange' 
collisions, the Cauchy and boundary conditions turn out to be  more than the
parameters in the trajectories, these are constrained by the following
condition
\be
\v\vi\v=[\v\xa\v+r_B+(n-1)\v\rb-\ra\v]/n\t.
\lb{c13}
\ee

The discussion about the existence and multiplicity of stationary points for
the action in the case of fixed Cauchy data for a particle initially
moving towards the wormhole mouth A, self-interacting via `collinear 
exchange' of velocities in Region II and traversing the wormhole 
$n$ times is essentially similar to that
done in the case of a `mirror exchange' type of collision.
In particular, for each fixed $n$, the Cauchy initial value problem has
a unique, globally self-consistent 
solution, given by Eqs. \rff{c12} and \rff{c10},
 provided that the initial Cauchy data are chosen to
satisfy the constraint \rff{c13} (and, of course, the kinematic
condition $\v\vi\v \leq c$).
Again the initial velocity $\v\vi\v$ is a monotonic decreasing 
function of $n$ in Eq. \rff{c13}, with extrema
\be
Min[c, [\v\xa\v+r_B]/\t]> \v\vi\v>\v\rb-\ra\v/\t.
\lb{c13a}
\ee

When the initial Cauchy data are fixed such that constraint \rff{c13a}
is satisfied, then Eq. \rff{c13} is solvable and it 
can be interpreted, for arbitrary and varying 
$n$, as fixing one of the coordinates of the collision point relatively
to the fixed wormhole mouths.

For example, if
\be
\t\geq {r_B+\v\xa\v\over c},
\lb{c16}
\ee
the kinematical condition $\v\vi\v<c$
is automatically satisfied independently
of the number $n$ of wormhole traversals. 
However, in the limit that $n\rightarrow \infty$, one sees from 
Eq. \rff{c13} that $\v\xa\v$ and, consequently, $\v\xi-\xa\v$ (for $\xi$
fixed in the frame \rff{c4}), become infinitely large.
This implies that, using arguments similar to those given for case {\bf a)}
(`mirror exchange' or `velocity exchange' collisions),
if the constraints \rff{c13a} and \rff{c16} are satisfied,
the multiplicity of the trajectory for `collinear exchange' motion
is in general finite, unless
the particle starts its motion in a region infinitely far from the
wormhole, in which case the multiplicity becomes infinite. 
Finally, the multiplicity is zero if condition \rff{c13a} is not
satisfied.
The lower bound on the initial velocity coming from condition \rff{c13a} 
is the same as conditions
\rff{me5} and \rff{75a}, which lead to infinitely many solutions
in cases {\bf a)} and {\bf b)}.
Again, the kinematical condition $\v\vi\v<c$ limits the set of
initial data whose trajectories have infinite multiplicity.\footnote{The
case of `collinear exchange' of velocities was not considered in Ref.
\cite{ekt}.
Assuming their ansatz for the initial data (now it is enough to give $\vi$,
while the information about $h$ is redundant), one can show that $\to$
is undetermined (function of the arbitrary $\xi$) and that multiplicity
is infinite provided that condition \rff{c13a} holds.}

$\bullet$ {\it `No collision'}

The other possible motion when $\vi$ is pointing towards mouth A is that for
which the particle directly enters mouth A and finally exits from mouth B
(after $n$ wormhole traversals), travelling forward in time and without
experiencing any self-collision.
The analysis of the kinematics for this motion is similar to that done
for the `no collision' case {\bf b)}.

In particular, using the following relation between the time of first
entrance into wormhole mouth A ($\tbn$) and the time of last exit
from mouth B ($\tbuno +\t$)
\be
\tbuno=\tbn+(n-1)\left[{\v\rb-\ra\v\over \v\vi\v}+\t\right ],
\lb{k1}
\ee
and imposing the boundary conditions (see also the footnote on pag. 6)
\be
\left. \begin{array}{l} 
\runo(\tbn)=\ra,
\\[0.5em]
\rnpuno(\tbuno+\t)=\rb
       \end{array}
\right.
\lb{k2}
\ee
and the usual Cauchy initial data \rff{unoa}, one can easily find that the 
trajectories are given again by the `no collision' formulas \rff{70b},
provided one interchanges $\rb$ with $\ra$ everywhere.
Moreover, one also finds for the time $\tbuno$
\be
\tbuno= \ti +{\v\ra-\ri\v\over \v\vi\v}+(n-1)\left[{\v\rb-\ra\v\over \v\vi\v}
+\t\right ],
\lb{k3}
\ee
for which, clearly, the second of conditions \rff{duea} for the 
existence of the time machine is trivially satisfied.

Therefore, for each fixed number $n$ of wormhole traversals, the Cauchy 
initial value problem for a particle initially heading towards mouth A
and exiting from mouth B without having self-collision has a unique
and globally self-consistent solution.
Since there are no constraints restricting the initial Cauchy data 
(apart from the
requirement $\v\vi\v<c$), the multiplicity of this set of trajectories
is always infinite.
The choice of initial Cauchy data made as in Ref. \cite{ekt} would again
formally lead to the same Eqs. \rff{70b} for the trajectories, but with
$\tbuno$ undetermined, and to infinite multiplicity of solutions even
for one single wormhole traversal.

In conclusion, the Cauchy initial value problem for the
case of a particle initially heading towards mouth A is classically ill-posed
(the multiplicity of each classical trajectory is always infinite).

\paragraph{ii) Ultrarelativistic velocity I/II: 
$\gamma_i \rightarrow \infty ~~;~~\gamma_k\sim finite ~~,~~k\not =i ~/~ 
\gamma_k \rightarrow \infty ~~,~~k=i, ... n+1$}
\hfill\break
Also in this case (with equations of the previous paragraph still
formally valid, modulo the substitution
 $\vec{v}_k\rightarrow c\vec{k}_k, \v\vec{k}_k\v=1$, for 
$k=1, ... n+1$) the multiplicity of trajectories is always infinite
and the Cauchy initial value problem is ill-posed.

\end{document}